\documentclass{ieeeaccess}
\usepackage{cite}
\usepackage{amsmath,amssymb,amsfonts,bm}
\usepackage{algorithmic}
\usepackage{graphicx}
\usepackage{textcomp}

\usepackage{comment}
\usepackage[framed,numbered,autolinebreaks,useliterate]{mcode} 
\usepackage{array,multirow} 
\usepackage{slashbox}

\def\BibTeX{{\rm B\kern-.05em{\sc i\kern-.025em b}\kern-.08em
    T\kern-.1667em\lower.7ex\hbox{E}\kern-.125emX}}
    \definecolor{mygreen}{RGB}{28,172,0} 
\definecolor{mylilas}{RGB}{170,55,241}

\newcommand{\R}{\mathbb{R}}

\newcommand{\bx}{\mathbf{x}}
\newcommand{\bc}{\mathbf{c}}

\newcommand{\bh}{\mathbf{h}}
\newcommand{\bd}{\mathbf{d}}

\newcommand{\bmm}{\mathbf{m}}
\newcommand{\bu}{\mathbf{u}}

\newcommand{\br}{\mathbf{r}}

\newcommand{\bv}{\mathbf{v}}

\newcommand{\bX}{\mathbf{X}}
\newcommand{\bY}{\mathbf{Y}}
\newcommand{\by}{\mathbf{y}}

\newcommand{\bA}{\mathbf{A}}

\newcommand{\bC}{\mathbf{C}}

\newcommand{\bH}{\mathbf{H}}
\newcommand{\bL}{\mathbf{L}}
\newcommand{\bM}{\mathbf{M}}
\newcommand{\bb}{\mathbf{b}}

\newcommand{\bS}{\mathbf{S}}

\newcommand{\bU}{\mathbf{U}}
\newcommand{\bT}{\mathbf{T}}
\newcommand{\bV}{\mathbf{V}}
\newcommand{\bI}{\mathbf{I}}
\newcommand{\bO}{\mathbf{0}}

\newcommand{\eqdef}{\overset{\mathrm{def}}{=\joinrel=}}
\newcommand\defeq{\stackrel{\text{\tiny def}}{=}}

\begin{document}
\history{Date of publication xxxx 00, 0000, date of current version xxxx 00, 0000.}
\doi{10.1109/ACCESS.2023.0322000}

\title{A new formula for faster computation of the k-fold  cross-validation and good regularisation parameter values in Ridge Regression}
\author{\uppercase{Kristian Hovde Liland}\authorrefmark{1},
\uppercase{Joakim Skogholt}\authorrefmark{1}, and \uppercase{Ulf Geir Indahl}\authorrefmark{1}}

\address[1]{ Faculty of Science and Technology, Norwegian University of Life Sciences, {\AA}s, Norway}
\tfootnote{The Research Council of Norway (project number 239070) provided financial support for this work.}

\markboth
{Liland \headeretal: Segmented
cross-validation residuals in linear regression}
{Liland \headeretal: Segmented
cross-validation residuals in linear regression}

\corresp{Corresponding author: Kristian Hovde Liland (e-mail: kristian.liland@nmbu.no).}

\begin{abstract}
In the present paper, we prove a new theorem, resulting in an update formula for linear regression model residuals calculating the exact k-fold cross-validation residuals for any choice of cross-validation strategy without model refitting. The required matrix inversions are limited by the cross-validation segment sizes and can be executed with high efficiency in parallel. The well-known formula for leave-one-out cross-validation follows as a special case of the theorem. 
In situations where the cross-validation segments consist of small groups of repeated measurements, we suggest a heuristic strategy for fast serial approximations of the cross-validated residuals and associated {\color{black}Predicted Residual Sum of Squares ($PRESS$)} statistic. We also suggest strategies for efficient estimation of the minimum $PRESS$ value and full $PRESS$ function over a selected interval of regularisation values.
The computational effectiveness of the parameter selection for  Ridge- and Tikhonov regression modelling resulting from our theoretical findings and heuristic arguments is demonstrated in several applications with real and highly multivariate datasets.
\end{abstract}


\titlepgskip=-21pt

\maketitle

\section{Introduction}
\label{sec:introduction}
\PARstart{M}{odel}-/parameter selection in statistical modelling is frequently justified from the maximum likelihood (ML) principle in combination with some measure of model quality (such as the Akaike's Information Criterion (AIC), Bayesian Information Criterion (BIC), Mallows $C_p$, the $PRESS$ statistic, etc.) that estimates the expected predictive performance for some candidate model(s) \cite{hastie09}.

According to Hjorth \cite{hjort93} the application of cross-validation measures as a methodology for model-/parameter selection in statistical applications was introduced by Stone \cite{stone74}. Stone's ideas motivated the invention of the generalised cross-validation ($GCV$) method by Golub et al. \cite{golub79} which is a computationally efficient approximation to the leave-one-out cross-validation (LooCV) method. It is invariant under orthogonal transformations and is considered to be a computationally efficient method for choosing appropriate regularisation parameter values in ridge regression (RR) modelling. {\color{black} Cross-validation is still an active area of research, see, e.g., \cite{benkeser2020improved,bates2023cross,smith2014correcting} for some recent works regarding prediction estimates for cross-validation, and \cite{celisse2016stability} for an analysis of the stability of generalisation bounds for leave-one-out cross-validation. The focus of the present work is the selection of an appropriate regularisation parameter value, primarily by grid search but numerical optimisation is also discussed.}

The RR method was introduced to the statistics community by Hoerl and Kennard \cite{hoerl70}, and is perhaps the most important special case in the Tikhonov  regularisation\cite{tikhonov63} (TR) framework of linear regression methods. The TR ideas were originally introduced to the community of numerical mathematics for solving linear discrete ill-posed problems in the context of inverse modelling. A good elementary introduction to the field is given in Hansen 
\cite{hansen10}.

The fast and exact calculations of the LooCV based {\it Predicted Residual Sum of Squares} ($PRESS$) statistic for the ordinary least squares (OLS) regression have been demonstrated by Allen \cite{allen71,allen74}. The purpose of the present paper is to demonstrate that such calculations are also available for the regularisation parameter selection problem of TR/RR at essentially no additional computational cost. In the present paper, we demonstrate this as follows: 

\begin{enumerate}
\item[i)]  From the Sherman--Morrison--Woodbury updating formula for matrix inversion, see Householder \cite{householder65}, we prove a new theorem that gives the general formula for calculating the segmented cross-validation (SegCV) residuals of linear least squares regression modelling. The formula for calculating the LooCV residuals in Allen's $PRESS$ statistic \cite{allen71,allen74} follows as a corollary of this result.

\item[ii)] We demonstrate how to obtain simple and fast LooCV calculations utilising the compact singular value decomposition (SVD) of a data matrix to quickly obtain $PRESS$ values associated with any choice of the regularisation parameter for a TR-problem.  
In particular, this enables fast graphing of the $PRESS$-values as a function of the regularisation parameter at any desired level of detail.

\item[iii)] For situations where some segmented cross-validation approach is required for obtaining the relevant $PRESS$-statistic values in the regularisation parameter selection, one may experience that even the segmented cross-validation formula from our theorem becomes computationally slow. To handle such situations, we propose an approximation of the segmented ($K$-fold) cross-validation strategy by invoking the computationally inexpensive LooCV strategy after conducting an appropriate orthogonal transformation of the data matrix. The particular orthogonal transformation is constructed from the left singular vectors of the $K$ local SVDs associated with each of the $K$ distinct cross-validation segments.\\ 
We demonstrate that the latter alternative provides practically useful approximations of the $PRESS$-statistic at substantial computational savings -- in particular for large datasets with many cross-validation segments (large $K$) containing either identical or highly related measurement values.
\end{enumerate} 

\section{Mathematical preliminaries}
\label{mathprelims}
If not otherwise stated we assume that $\bX$ is a centred $(n\times p)$ data matrix ($\bX'$ denotes the transpose of $\bX$) and that the corresponding $(n\times 1)$ vector $\by$ of responses is also centred. We define the scalar $\bar{y}$ and row vector $\bar{\bx}$ as the (column) averages of ${\by}$ and ${\bX}$ obtained before centring, respectively.

\subsection{Model estimation in ordinary least squares and ridge regression}
In ordinary least squares (OLS) regression \cite{hastie09} one minimises the {\it residual sum of squares}
\begin{equation}\label{OLS}RSS(\bb)=\|\bX\bb-\by\|^2,\end{equation}
to identify the least squares solution(s) of (\ref{OLS}) with respect to the regression coefficients ${\bb}$. A least squares solution $\bb_{OLS}$ of (\ref{OLS}) corresponds to an exact solution of the associated {\it normal equations}
\begin{equation}\label{NormEqs}\bX'\bX\bb=\bX'\by,\end{equation}
where $\bb_{OLS}$ is unique when $\bX'\bX$ is non-singular. For later predictions of uncentred data, the associated vector of fitted values is given by 
\begin{equation}\label{FittedVals}\hat{\by}=\bX\bb_{OLS}+b_0,\end{equation} 
where the constant term (intercept) $b_0 = \bar{y}-\bar{\bx}\bb_{OLS}$.

{\color{black}
For centred vectors/matrices, $\by$ and $\bX$, this equation becomes $\hat{\by}=\bX\bb_{OLS}=\bH\by$. Here, the projection matrix, $\bH$, (a.k.a. the hat matrix) is defined as
\begin{equation}\label{ProjMat}\bH\defeq\bX(\bX^\prime\bX)^{-1}\bX^\prime=\bT\bT^\prime, \end{equation} 
where $\bT$ can be chosen as any orthogonal $(n\times r)$-matrix spanning the column space of the centred $\bX$-data.}

For various reasons a minimiser $\bb_{OLS}$ of $RSS(\bb)$ in equation (\ref{OLS}) is not always the most attractive choice from a predictive point of view \cite{hastie09,hansen10,kalivas12}. For instance ${\bX}'{\bX}$ may be singular or poorly conditioned, the solution of (\ref{NormEqs}) is not unique or inappropriate etc. An alternative and quite useful solution was independently recognised by Tikhonov \cite{tikhonov63}, Phillips \cite{phillips62}, and Hoerl and Kennard \cite{hoerl70}. Instead of directly minimising $RSS(\bb)$, their alternative proposal was to minimise the weighted bi-objective least squares problem
\begin{equation}\label{RR}
RSS_\lambda(\bb)=\|\bX\bb-\by\|^2+\lambda\|\bI\bb-{\mathbf 0}\|^2
=\|\bX\bb-\by\|^2+\lambda\|\bb\|^2,
\end{equation}
where the scalar $\lambda>0$ is a fixed {\it regularisation parameter} (of appropriate magnitude), the matrix $\bI$ is the $(p\times p)$ identity matrix and ${\mathbf 0}$ is a $(p\times 1)$ vector of zeros.
This formulation explicitly represents a penalisation with respect to the Euclidean $(L_2)$ norm $\|\bb\|$ of the regression coefficients. {\color{black}The identity matrix $\bI$ can also be replaced by an alternative regularisation matrix $\bL$ as described in Appendix \ref{secL2}.} For a fixed $\lambda$, the unique minimiser of (\ref{RR}) is given by  $\bb_\lambda$ of equation (\ref{bRR1}) below. The rightmost part of Equation (\ref{RR}) is sometimes referred to as a TR-problem in {\it standard form} \cite{hansen10}. 

The minimisation of equation $(\ref{RR})$ with respect to $\bb$ is  equivalent to solving the OLS problem associated with the augmented
data matrix and response vector:
\begin{equation}\label{Xlambda}\bX_{\lambda}=\left[%
\begin{array}{c}
  \bX \\
  \sqrt{\lambda}\bI \\
\end{array}%
\right], \ \ \by_0=\left[%
\begin{array}{c}
  \by \\
{\mathbf 0} \\
\end{array}%
\right].\end{equation}
Note that linear independence of the $\bX_\lambda$-columns trivially follows from linear independence of the $\bI$-columns.
The matrix product $\bX_{\lambda}^\prime\bX_{\lambda}$ in the associated normal equations 
\begin{equation}\label{NormalRR1}\bX_{\lambda}^\prime\bX_{\lambda}\bb=\bX_{\lambda}^\prime\by_0\end{equation}
is therefore non-singular, and the corresponding least squares solution 
\begin{equation}\label{bRR1}\bb_{\lambda}=(\bX_{\lambda}^\prime\bX_{\lambda})^{-1}\bX_{\lambda}^\prime\by_0\end{equation}
of the augmented problem \eqref{Xlambda} becomes unique.
	Straight forward algebraic simplifications of (\ref{NormalRR1}) result in the 
the familiar normal equations associated with the RR-problem
\begin{equation}\label{normalRR2}({\bX}^\prime{\bX}+\lambda{\bI}){\bb}={\bX}^\prime{\by},\end{equation}
and the solution in (\ref{bRR1}) simplifies to
\begin{equation}\label{bRR2}{\bb}_{\lambda}=({\bX}^\prime{\bX}+\lambda{\bI})^{-1}{\bX}^\prime{\by}.\end{equation}
For subsequent applications of the $\lambda$-regularised model to uncentred $\bX$-data,  
the appropriate constant term in the resulting regression model is
\begin{equation} \label{constantTerm}
b_{0,\lambda}=\bar{ y}-\bar{\bx}{\bb}_{\lambda},
\end{equation} and the associated vector of fitted values $\hat{\by}_{\lambda}$ is given by
\begin{equation}\label{FittedValsRR}\hat{\by}_{\lambda}={\bX \bb}_{\lambda}+b_{0,\lambda}.\end{equation}

{\color{black} The full SVD of $\bX = \bU\bS\bV^\prime$ yields $\bV\bV^\prime =\bI_p$ and $\bX^\prime\bX = \bV\bS^\prime\bS\bV^\prime$. Assuming that $\bX$ has full rank it is shown in Appendix \ref{Simplifications} that the regression coefficients are given by $\bb_{\lambda} = \bV_r\bc_\lambda$, where $\bV_r$ are the right singular vectors of the compact SVD and the coordinate vector
$\bc_\lambda$ has the scalar entries 
\begin{equation}\label{ci}
c_{\lambda,j} = \frac{\bu_j^\prime\by}{s_j+\lambda/s_j}, \text{ for } 1\leq j\leq r. 
\end{equation}
Compared to the relatively large computational costs associated with calculating the (compact) SVD of $\bX$, calculation of the regression coefficient candidates (even for a large number of different $\lambda$-values) only requires computing the vectors $\bc_\lambda$ according to Equation (\ref{ci}) and the matrix-vector multiplications $\bb_\lambda=\bV_r\bc_\lambda$ as derived in Equation (\ref{bRRSVD}).

For the regularised multivariate regression with several $(q)$ responses, ${\bY}\in\mathbb{R}^{n\times q}$,  the associated matrix of regression coefficients is
\begin{equation} \label{BRRSVD}
[\bb_{1,\lambda}\ ...\ \bb_{q,\lambda}] = \bV_r(\bS_r+\lambda\bS_r^{-1})^{-1}\bU_r^\prime\bY=\bV_r\bC_\lambda,
\end{equation} 
where $\bC_\lambda=(\bS_r+\lambda\bS_r^{-1})^{-1}\bU_r^\prime\bY$ is the obvious multivariate generalisation of the vector $\bc_\lambda$ introduced above.
}

\subsection{Obtaining cross-validation segments by projection matrix correction}
\label{presstheory}
When the columns of the data matrix $\bX$ are linearly independent, the associated OLS-solution $\bb_{OLS}$ of the normal equations (\ref{NormEqs}) is unique, and cross-validation residuals can be derived from the Sherman--Morrison--Woodbury formula for updating matrix inverses \cite{householder65}. From Theorem \ref{TmSegCV} in the Appendix, we obtain the general segmented CV (SegCV) residuals
\begin{equation}\label{eq:seg_residuals}
\br_{(\{k\})} =  [\bI_{n_k} - 
\bH_{\{k\}}]^{-1} \br_{\{k\}},
\end{equation}
where $\{k\}$ refers to the samples of the $k$-th CV segment, 
$\br_{(\{k\})}$
refers to the vector of predicted residuals when the segment samples are not included in the modelling, $n_k$ is the number of samples in the segment and, $\bH_{\{k\}}$ is the sub-matrix of the projection matrix $\bH$ (defined in Equation (\ref{ProjMat}) above) associated with the samples of the $k$-th CV segment. This means that updating residuals for a given segment entails the inversion of a matrix involving the entries of $\bH$  corresponding to all pairs of sample indices of the $k$-th CV segment. The computational cost of the inversions obviously depends on the number of segments and the number of samples belonging to each segment. 

{\color{black}Allen \cite{allen71,allen74} suggested the $PRESS$ (Prediction Sum-Of-Squares) statistic
\begin{equation}\label{PRESS_Allen}
PRESS=\sum_{i=1}^n(y_i-\hat{y}_{i,(i)})^2 = \sum_{i=1}^n \br_{(i)}'\br_{(i)}.
\end{equation}
\noindent where $\hat{y}_{i,(i)}$ denotes the OLS prediction of the $i$-th sample when the sample has been deleted from the regression estimation, and $\br_{(i)}$ is the corresponding residual. With $\hat{y}_{i,(\{k\})}$ denoting the predictions of the $i$-th sample after deleting the corresponding $k$-th CV segment  samples from the regression problem in \eqref{OLS}, the SegCV equivalent of the $PRESS$-statistic becomes}:

\begin{equation}\label{PRESS_segmented}
\begin{split}
PRESS &=\sum_{i=1}^n(y_i-\hat{y}_{i,(\{k\})})^2 = \sum_{k=1}^K \br_{(\{k\})}'\br_{(\{k\})} \\
&=\sum_{k=1}^K \sum_{i=1}^{n_k} r_{i,\{k\}}^2.
\end{split}
\end{equation}

\noindent Here $r_{i,\{k\}}$ are the elements of the residual vectors defined in Equation \ref{eq:seg_residuals}.

\subsubsection{The leave-one-out cross-validation} 
Corollary \ref{CoLooCV} of Theorem \ref{TmSegCV} covers the special case of LooCV where Equation
(\ref{eq:seg_residuals})
simplifies to a computationally efficient scalar formula for updating the individual residuals
\begin{equation}
 r_{(i)} = r_i/(1-h_i).
\end{equation}


{\color{black}$h_i$ is often referred to as the {\it leverage value} associated with the $i$-th sample (row) in $\bX$.} For $\hat{y}_{i,(i)}$ denoting the prediction of the $i$-th sample after deleting it from the regression modelling problem in \eqref{OLS}, the LooCV $PRESS$-statistic, is given by

\begin{equation}\label{PRESS}PRESS=\sum_{i=1}^n(y_i-\hat{y}_{i,(i)})^2
=\sum_{i=1}^n\left(\frac{y_i-\hat{y}_{i}}{1-h_i-1/n}\right)^2.\end{equation}

\noindent
In (\ref{PRESS}) $\hat{y}_{i}$ is the $i$-th entry in the vector of fitted values $\hat{\by}=\bX\bb_{OLS}+b_0$, and $h_i$ denotes the $i$-th diagonal element  of the projection matrix $\bH$ defined in (\ref{ProjMat}) above. The denominator $(1-h_i-1/n)$ scales the $i$-th model residual $(y_i-\hat{y}_i)$ to obtain the exact LooCV prediction residual $(y_i-\hat{y}_{i,(i)})$. The term $1/n$ in this denominator accounts for the centring of the $\bX$-columns and the associated inclusion of a constant term $(b_0)$ in the regression model (\ref{FittedVals}).

From the last identity in Equation (\ref{ProjMat}) it is clear that the entries of the $n$-vector $\bh = [h_1\ h_2\ ... \ h_n]^\prime$, corresponding to the diagonal elements of $\bH$, are identical to the square of the norms of the $\bT$-rows, i.e.
\begin{equation}\label{h_leverages}
\bh=(\bT\odot\bT){\mathbf 1}.
\end{equation}
Here, $\bT\odot\bT$ denotes the Hadamard (element-wise) product of $\bT$ with itself and ${\mathbf 1}\in \mathbb{R}^r$ is the constant vector with $1$'s in all entries. Appropriate choices of the matrix $\bT$ can be obtained in various ways including both the QR-factorisation and the SVD of $\bX$. 

It should be noted that calculating the matrix inverse $(\bX^\prime\bX)^{-1}$ in the process for finding the diagonal $\bh$ of $\bH$ in (\ref{ProjMat}) is neither required nor recommended in practice. In general, the explicit calculation of matrix inverses (for non-diagonal matrices) should be avoided whenever possible due to various unfavourable computational aspects, see Bj{\"o}rck \cite[Section 1.2.6]{bjorck16}. 

\subsubsection{The generalised cross-validation} 
The $GCV(\lambda)$ was proposed by Golub et al. \cite{golub79} as a fast method for choosing good regularisation parameter ($\lambda$) values in RR. Here, we consider the definition  
\begin{equation}\label{GCVdef}
\begin{split}
GCV(\lambda) &\eqdef \sum_{i=1}^n\left(\frac{y_i-\hat{y}_{\lambda, i}}{1-\bar{h}_\lambda-1/n}\right)^2 \\
&=(1-df(\lambda)/n)^{-2}\|\by-\bX\bb_\lambda\|^2,
\end{split}
\end{equation}
where $(y_i-\hat{y}_{\lambda, i})$ is the $i$-th entry of the residual vector $\br_\lambda=\by-\hat{\by}_\lambda$, $\bar{h}_\lambda\defeq\frac{1}{n}\sum_{j=1}^{r}\frac{s_j}{s_j+\lambda/s_j}$ and the effective degrees of freedom $df(\lambda)\eqdef n\bar{h}_\lambda+1$. This definition of $GCV(\lambda)$ is proportional (by the sample size $n$) to the definition given in \cite[page 216]{golub79}. The $GCV(\lambda)$ is explained as a rotation invariant alternative to the LooCV that provides an approximation of the $PRESS(\lambda)$-statistic defined below.

From the elementary matrix-vector multiplication formula (\ref{bRRSVD}) for computing the regression coefficients $\bb_\lambda$, it is clear that $GCV(\lambda)$ can be calculated very efficiently for a large number of different $\lambda$-values once the 
non-zero singular values of $\bX$ are available.

In their justification of $GCV(\lambda)$ as the preferable choice over the exact LooCV-based $PRESS(\lambda)$, Golub and co-workers stressed the unsatisfactory properties of the $PRESS$-function when the rows of $\bX$ are exactly or approximately orthogonal. In this case, the estimated regression coefficient $\bb_\lambda^{(i)}$ (obtained by excluding the $i$-th row $\bx_i$ of $\bX$) must be correspondingly orthogonal (or nearly orthogonal) to the excluded sample $\bx_i$. Consequently, the associated leave-one-out prediction $\hat{y}_{i,(i)}(=\bx_i\bb_\lambda^{(i)})$ becomes a poor estimate of the corresponding $i$-th response value $y_i$. 

NOTE: In situations such as the one just described, it makes little sense to think of the $\bX$-data as a collection of independent random samples, and the statistical motivation for considering the LooCV idea becomes correspondingly inferior. 
{\color{black}In \cite{golub79} it is claimed that any parameter selection procedure should be invariant under orthogonal transformations of the $(\bX,\by)$-data. We are sceptical of this requirement as an inexpedient restriction. This relates to the context of approximating the $PRESS$-statistic for situations where a segmented/folded cross-validation approach is appropriate.}

\section{Calculation of the cross-validation based $PRESS(\lambda)$-functions} 

From Equations (\ref{PRESS_segmented}, \ref{PRESS}) and the matrix- and vector augmentations in Equation (\ref{Xlambda}), it is clear that the computationally fast versions of the SegCV and LooCV with the associated $PRESS$-statistic are also valid for TR-problems when the regularisation parameter $\lambda$ is treated as a fixed quantity. 

Below we will first handle the general case of segmented cross-validation. Thereafter we derive an equation assuring fast calculations of the regularised leverages in the vectors $\bh_\lambda$ necessary for the LooCV situation. The required calculations are remarkably similar to a computationally efficient calculation of the fitted values $\hat{\by}_\lambda$
and closely related to the corresponding regularised regression coefficients $\bb_\lambda$ in (\ref{bRRSVD}). Both $\bh_{\lambda}$, $\hat{\by}_\lambda$ (and $\bb_\lambda$) can be obtained from the SVD of the original centred data matrix $\bX$. This makes the computations of the exact LooCV-based $PRESS(\lambda)$-function defined in (\ref{PRESSlambda}) below about as efficient as the approximation obtained by the $GCV(\lambda)$ in (\ref{GCVdef}).

\subsection{Exact $PRESS(\lambda)$-functions from the SVD of the augmented matrix $\bX_\lambda$}\label{SVD_alt}
Again, we assume that the centred $\bX$ has full rank $r$ and that $\bX=\bU_r\bS_r\bV_r^\prime$ is the associated compact SVD. By defining $\bS_{\lambda,r}$ to be the diagonal $r\times r$ matrix with non-zero diagonal entries $\sqrt{s_j^2+\lambda}, \ j=1,...,r$, the $r$ most dominant singular values of the augmented matrix $\bX_{\lambda}$ in (\ref{Xlambda}) are given by the diagonal elements of $\bS_{\lambda,r}$. From equation (\ref{XtXlambdaI}) in Section \ref{mathprelims}, the right singular vectors $\bV_r$ of $\bX$ are also the right singular vectors of $\bX_\lambda$, and the associated $r$ left singular vectors are given by
\begin{equation}\label{Ulambda}
\begin{split}
\bT_{\lambda,r} &= 
\bX_{\lambda}\bV_r\bS_{\lambda,r}^{-1}=\left[%
\begin{array}{c}
  \bX\bV_r\bS_{\lambda,r}^{-1} \\
  \sqrt{\lambda}\bI\bV_r\bS_{\lambda,r}^{-1} \\
\end{array}%
\right] \\
&=\left[%
\begin{array}{c}
  \bU_r\bS_r\bS_{\lambda,r}^{-1} \\
  \sqrt{\lambda}\bV_r\bS_{\lambda,r}^{-1} \\
\end{array}%
\right]=\left[%
\begin{array}{c}
\bU_{\lambda,r} \\
  \sqrt{\lambda}\bV_r\bS_{\lambda,r}^{-1} \\
\end{array}%
\right],
\end{split}
\end{equation}
where the matrix $\bU_{\lambda,r}\defeq \bU_r\bS_r\bS_{\lambda,r}^{-1}$ denoting the upper $n$ rows of $\bT_{\lambda,r}$ is the part of actual interest (the additional left singular vectors not included in (\ref{Ulambda}) are all zeros in the upper $n$ entries). Because $\bS_r\bS_{\lambda,r}^{-1}$ is $(r\times r)$ diagonal, $\bU_{\lambda,r}$ is obtained by scaling the $j$-th column $(1\leq j\leq r)$ of 
$\bU_r$ with $\sqrt{s_j/(s_j+\lambda/s_j)}$. 

From the above definition of $\bU_{\lambda,r}$, calculation of the $PRESS$-residuals associated with the $n$ original $(\bX,\by)$ data points in the augmented least squares problem $\bX_\lambda\bb=\by_0$ is straight forward. According to Equations (\ref{ProjMat}, \ref{Ulambda}), the regularised hat matrix $\bH_\lambda$ is given by 
\begin{equation}\label{hlambdaSeg}\bH_\lambda= \bU_{\lambda,r} \bU^\prime_{\lambda,r}
.
\end{equation} 

For each choice of the regularisation parameter $\lambda>0$ and the corresponding expression for the regression coefficients $\bb_\lambda$ in Equation (\ref{bRRSVD}), the fitted values are \begin{equation}\label{yhat}\hat{\by}_\lambda=\bX\bb_\lambda+b_{0,\lambda}=(\bU_r\bS_r)\bc_\lambda+b_{0,\lambda}\end{equation} $$= \bH_\lambda\by+b_{0,\lambda}.$$ Hence, 
\begin{equation}\label{PRESSlambdaSeg}PRESS(\lambda)\eqdef \sum_{k=1}^K \| [\bI_{n_k} - \bH_{\lambda,\{k\}}-1/n]^{-1} (\by_{\{k\}}-\hat{\by}_{\lambda,\{k\}})\|^2,\end{equation}
where $\by_{\{k\}}-\hat{\by}_{\lambda,\{k\}}$ is the sub-vector of the residual vector ${\br}_\lambda=\by-\hat{\by}_\lambda$ corresponding to the $k$-th CV segment and 
$\bH_{\lambda,\{k\}}$ is the associated sub-matrix of $\bH_\lambda$. 
While 
Equation (\ref{PRESSlambdaSeg}) defines the general, segmented cross-validation case, the special case of LooCV simplifies considerably. Only the diagonal entries of $\bH_\lambda$ (the sample leverages) are required, i.e., 
Equation (\ref{PRESSlambdaSeg}) 
simplifies to
\begin{equation}\label{PRESSlambda}PRESS(\lambda)\eqdef \sum_{i=1}^n\left(\frac{y_i-\hat{y}_{\lambda,i}}{1-h_{\lambda,i}-1/n}\right)^2.\end{equation}
Note that $\bar{h}_\lambda$ in the denominator of Equation (\ref{GCVdef}) defining $GCV(\lambda)$ is identical to the mean of the $\bh_\lambda$-entries, i.e. $\bar{h}_\lambda=(1/n)\sum_{i=1}^nh_{\lambda,i}$, due to the fact that $\bU_r$ is an orthogonal matrix.
Also note that the diagonal entries of $\bH_\lambda$ can be calculated directly by
\begin{equation}\label{hlambda}\bh_\lambda= (\bU_{\lambda,r}\odot \bU_{\lambda,r}){\mathbf 1}=(\bU_r\odot \bU_r)\bd_{\lambda}, \end{equation}
where the coefficient vector $\bd_{\lambda}=[d_{1,\lambda}\ ...\ d_{r,\lambda}]^\prime=(\bS_r\bS_{\lambda,r}^{-1})^2{\mathbf 1}\in \R^r$ has the entries 
\begin{equation}\label{di}
d_{i,\lambda} = \frac{s_j^2}{s_j^2+\lambda}=\frac{s_j}{s_j+\lambda/s_j},\text{ for } 1\leq j\leq r.
\end{equation}

Consequently, the evaluation of the $PRESS(\lambda)$-function defined in (\ref{PRESSlambda}) is essentially available at the additional computational cost of two matrix-vector multiplications (Equations (\ref{yhat},\ref{hlambda})) where the matrices ($\bU_r\bS_r$ and $\bU_r\odot\bU_r$) are fixed and the associated coefficient vectors 
$\bc_\lambda$ and $\bd_\lambda$
are obtained by elementary arithmetic operations for each choice of $\lambda>0$. 
{\color{black} A note on the number of floating point operations (flops) required for the fast calculation of the LooCV-based $PRESS(\lambda)$-function is included in Appendix \ref{floploocv}}. 

\subsection{Alternative strategies for estimating the SegCV-based $PRESS(\lambda)$-function}
The LooCV calculations in the previous section can be implemented at low computational costs dominated by the SVD of $\bX$. The SegCV version, however, also involves the inversion of several matrices associated with each combination of the regularisation parameter value of $\lambda$ and cross-validation segment. In situations with many CV segments, e.g., defined by relatively small groups of replicates, the additional computational costs may be acceptable as the matrices to be inverted are small. However, for large datasets with few segments, e.g., 5-10, the required amount of computations may be rather large (comparable to explicitly holding out samples and recalculating a full TR model from scratch for each CV segment).

We therefore describe two alternative strategies for speeding up calculations. The first one is based on approximating the $PRESS$-values, while the second strategy involves clever usage of a small subset of exact $PRESS(\lambda)$-values to estimate the minimum of the $PRESS(\lambda)$-value and/or the complete $PRESS(\lambda)$ curve within some range of the regularisation parameter value.

\subsubsection{$PRESS(\lambda)$ approximated by segmented virtual cross-validation -- VirCV}


We will consider a faster alternative for approximating the SegCV approach for the type of situations just described. In the following, we assume (without loss of generality) that the uncentred data matrix 
\begin{equation}\label{UnceteredForT}
\begin{split}
&\bX = \left[%
\begin{array}{c}
  \bX_1 \\
  \bX_2 \\
  : \\
  \bX_K
\end{array}%
\right]\text{together with the uncentred response vector }\\
&\by = \left[%
\begin{array}{c}
  \by_1 \\
  \by_2 \\
  : \\
  \by_K
\end{array}%
\right]\ (K\geq 2)
\end{split}
\end{equation}
is composed by $K$ distinct sample segments. For $1\leq k\leq K$, we assume that $\bU_k\bS_k\bV_k^\prime = \bX_k$ denotes the compact SVD of segment number $k$, and that $n_k$ is the number of rows in $\bX_k$ so that the total number of samples is $n = \sum_{k=1}^{K}n_k$. 

From the SVD of the $k$-th segment, we obtain the identity $\bU_k^\prime\bX_k = \bS_k\bV_k$. Consequently, the orthogonal transformation performed by left multiplication with the $(n_k\times n_k)$ matrix $\bU_k^\prime$ transforms the samples segment $\bX_k$ into a matrix of strictly orthogonal rows. 
Now we define the two block diagonal matrices
\begin{equation} \label{tdef}
\bT = \left[%
\begin{array}{cccc}
  \bU_1 & & & \\
   & \bU_2 & & \\
  & & \ddots & \\
  & & & \bU_K
\end{array}%
\right] \text{ and }
\tilde{\bT}=\left[\begin{array}{cccc} \bT & \bm{0} \\ \bm{0} & \bI\end{array}\right],
\end{equation}
with the properties $\bT^\prime\bT=\bT\bT^\prime=\bI$ and $\tilde{\bT}^\prime\tilde{\bT}=\tilde{\bT}\tilde{\bT}^\prime=\bI$, i.e., both $\bT$ and $\tilde{\bT}$ are orthogonal.

The formulation of TR-modelling for uncentred $\bX$ and explicit inclusion of the constant term corresponds to finding the least squares solution of the linear system

\begin{equation}\label{boeq1}
\left[\begin{array}{cc} \bm{1} & \bX \\ \bm{0} & \sqrt{\lambda}\bI\end{array}\right] \cdot \left[\begin{array}{c} b_0 \\ \bm{b}\end{array}\right] = \left[\begin{array}{c} \by \\ \bm{0} \end{array}\right],
\end{equation}
and left multiplication of \eqref{boeq1} by 
the orthogonal matrix $\tilde{\bT}^\prime$ yields the system

\begin{equation}\label{boeq2}
\left[\begin{array}{cc} \bT^\prime\bm{1} & \bT^\prime\bX \\ \bm{0} & \sqrt{\lambda}\bI\end{array}\right] \cdot \left[\begin{array}{c} b_0 \\ \bm{b}\end{array}\right] = \left[\begin{array}{c} \bT^\prime\by \\ \bm{0} \end{array}\right].
\end{equation}

Note that the associated normal equations of the systems in \eqref{boeq1} and \eqref{boeq2} are identical. Hence, their least squares solutions are also identical. 

\textbf{Definition of the segmented virtual cross-validation} \\
We define the \textit{segmented virtual cross-validation (VirCV)} strategy as the process of applying the LooCV strategy to the transformed system in equation \eqref{boeq2}. As is noted above, multiplication by $\bT^\prime$ has the effect of orthogonalising the rows within each of the $K$ segments in the $\bX$ matrix.

The heuristic argument for justifying the VirCV approach as an approximation of a SegCV 
approach is that the rows within each transformed data segment are unsupportive of each other under the LooCV strategy (due to the internal "decoupling" of each segment into a set of mutually orthogonal row vectors). However, from practical cases, it can be observed that the accuracy of this approximation depends on the level of similarity between the original samples within each segment of data points.

Note that contrary to the LooCV, the $GCV$ is not useful in combination with the VirCV strategy. The reason for this is that the singular values of $\bX$ are invariant under orthogonal transformations. From equation (\ref{GCVdef}) and the definition of $\bar{h}_\lambda$ it follows that $GCV(\lambda)$ is also invariant under orthogonal transformations, i.e., 
the systems in \eqref{boeq1} and \eqref{boeq2} lead to the same $GCV(\lambda)$-function .

With the VirCV we are clearly cross-validating on 
the orthogonal phenomena caused by the samples within each segment. 
As all the samples in a segment contribute to identifying these directions, the VirCV cannot be expected to provide exactly the same results as the SegCV. One may, however, expect that when the different segments are carefully arranged to contain highly similar samples only (which is a reasonable assumption to make for most organised studies with such data segments), then the VirCV should provide a useful approximation to the SegCV. This will be demonstrated in the application section below. For special situations deviating from highly similar samples in the segments, see Appendix \ref{app:VirCVsituations}.

\vspace{3mm}

\noindent
\textbf{Computational aspects in the leverage corrections for the VirCV}\\ As is noted in association with (\ref{UnceteredForT}), the VirCV procedure requires an initial calculation of the transformation $\bT$ from the segments of the uncentred $\bX$-data. For a correct implementation of the computational shortcuts similar to those of the LooCV, it is necessary to mean centre the data matrix $\bX$ prior to executing the $\bT$-transformation and the least squares modelling. In practice, one must therefore mean centre the data prior to the multiplication with $\bT^\prime$ (or, equivalently, one can multiply by $\bT^\prime$ and subtract the projection of the transformed data onto the transformed vector $\bT^\prime\bm{1}$ of ones). As $\bT$ is an orthogonal transformation the angles and in particular the orthogonality between vectors will be preserved.
For the transformed data, modelling by including a constant term is therefore associated with the transformed vector
$\bT^\prime\bm{1}$ of ones. With $\bX_c$ and $\by_c$ denoting the centred data matrix and the associated centred response vector, respectively, the vector $\bT^\prime\bm{1}$ is orthogonal to the columns of the transformed centred data $\bT^\prime\bX_c$ and $\|\bT^\prime\bm{1}\|=\|\bm{1}\|=\sqrt{n}$. The justification for the leverage correction described earlier therefore still holds, but the particular correction terms ($1/n$) changes. 

{\color{black}
With the transformed centred predictors $\tilde{\bX}=\bT^\prime\bX_c$ and responses $\tilde{\by}=\bT^\prime\by_c$ in \eqref{boeq2}, the associated fitted values as $\hat{\tilde{\by}}_\lambda=\tilde{\bX}\bb_\lambda$},
the $PRESS$-function for the VirCV is given by
\begin{equation}\label{PRESSVirCV}
\begin{split}
PRESS_{VirCV}(\lambda) &=\sum_{i=1}^n(\tilde{{y}}_i-\hat{\tilde{{y}}}_{\lambda,i,-1})^2 \\
&=\sum_{i=1}^n\left(\frac{\tilde{{y}}_i-\hat{\tilde{{y}}}_{\lambda,i}}{1-h_{\lambda,i}-m_i/n}\right)^2.
\end{split}
\end{equation}
Here the leverages $h_{\lambda,i}$ are calculated as in (\ref{hlambda}) based on the transformed version $\tilde{\bX}$ of the centred data, and the enumerator of the correction terms are the entries of the vector $\bmm=\bT^\prime\bm{1}\odot\bT^\prime\bm{1}\in\mathbb{R}^n$. This means that the correction term $1/n$ in the denominator of (\ref{PRESSlambda}) must be replaced by $m_i/n$ in (\ref{PRESSVirCV}), where $m_i$ denotes the $i$-th entry of the vector $\bmm$ (to be consistent with the orthogonal transformation of the regularised least squares problem).

{\color{black}A comparison of the number of flops required for the VirCV compared to the SegCV is included in Appendix \ref{flopVirCV}}. 

{\color{black}\subsubsection{Approximated $PRESS$-function using subsets of $\lambda$}
\textbf{Minimum $PRESS$-value estimation} \\
If TR is used in an automated system (without subjective assessment) or only the optimal $PRESS(\lambda)$ is needed, we can avoid redundant calculations by searching for the $\lambda$ value that minimises (\ref{PRESSlambdaSeg}) instead of calculating a large range of solutions. A possible approach for such a search can be based on the golden section search with parabolic interpolation \cite{brent1973algorithms}. This method performs a search for the minimal function value over a bounded interval of a single parameter. To leverage the previously described efficient computations of fitted values, $\hat{\by}_\lambda$, coefficient vectors, $\bd_\lambda$, etc. the search for minimum $PRESS(\lambda)$ is then performed over a fixed set of $\lambda$-values. The grid of $\lambda$-values can have high resolution while still achieving a considerable advantage in computational speed compared to the exhaustive $PRESS$-function calculations. It is well-known that this type of function minimisation cannot guarantee the optimal value to be found, however, the $PRESS$-functions of interest often have relatively smooth and simple graphs, where a global minimum over the $\lambda$-interval of interest can be found with high accuracy.

$PRESS(\lambda)$\textbf{-function estimation by spline interpolation} \\
In cases where estimating the detailed $PRESS(\lambda)$-function is beneficial, e.g., for plotting and inspection, it may be possible to reduce the number of accurate $PRESS(\lambda)$-evaluations to be calculated quite substantively without sacrificing much precision in the estimation. 

We propose a cubic spline strategy, where the $PRESS(\lambda)$-function is estimated from a small set of distinct $\lambda$-values, and new values are added to the set iteratively until the difference between estimation and true $PRESS$-value falls below a chosen threshold for all $\lambda$-values in the extended set. The latter is determined by cross-validation of the cubic spline interpolation, i.e., a low-cost operation. 

As with the $PRESS$-minimisation procedure, we consider a fixed set of $\lambda$-values from which we choose starting points and select subsequent values. The $\lambda$-values extending the set in each iteration are the ones halfway to neighbours of the chosen $\lambda$-values on both sides, effectively doubling the local density of $\lambda$-values where needed (low accuracy of spline approximation). Starting values for the initial set of $\lambda$s can be chosen equidistant (on a log$_{10}$ scale) or the sequence obtained using the above "Minimum $PRESS$-value estimation" strategy. Experience with real datasets indicates that the latter is an efficient strategy that may provide close to exact estimation of the minimum $PRESS$-value.}

\subsection{A short note on model selection heuristics}
With the key formulas derived above we obtain efficient model selection procedures from 
minimising the $PRESS(\lambda)$- or the $GCV(\lambda)$-functions with respect to the regularisation parameter $\lambda$. However, the minima of these functions will not necessarily assure the selection of the best model in terms of future predictions. This is particularly the case when the $PRESS$- and $GCV$ functions are relatively flat for a relatively large interval of $\lambda$s containing the minimum value. In such situations it is often useful to invoke heuristic principles such as {\it Occam's razor} for identifying a simpler model (in terms of the norm of the regression coefficients) at a small additional cost in terms of the $PRESS$ (or the $GCV$):

The '\textbf{1 standard error rule}' described in \cite{hastie09} obtains a simpler (more regularised) alternative by selecting a model where the $PRESS$-statistic is within one standard error of the $PRESS$-minimal model. More precisely, we first identify the minimum $PRESS$ value and calculate the standard error of the squared cross-validation errors associated with this model. Then the largest regularisation parameter value where the associated model has a $PRESS$-statistic within one standard error of the $PRESS$-minimum is selected. 

The '$\chi^2$ \textbf{model selection rule}' to determine the regularisation parameter
was originally introduced for model selection with Partial Least Squares regression modelling \cite{indahl05}. 
By assuming that the residuals associated with the minimum value $PRESS_{min}$ of $PRESS(\lambda)$ are randomly drawn from a normal distribution, the statistic given by $n\cdot PRESS_{min}/\sigma^2$, where $\sigma^2$ is the associated (unknown) variance, follows a $\chi_n^2$ distribution (where $n$ is the degrees of freedom). By fixing a particular significance level $\alpha$, 
the selection rule says: {\it Choose the largest possible value of $\lambda$ so that $n\cdot PRESS_{min}/PRESS(\lambda)\geq \chi_{n,\alpha}^2$}. Here, $\chi_{n,\alpha}^2$ is the lower $\alpha$-quantile of the $\chi_{n}^2$ distribution and $PRESS(\lambda)$ is a substitute for $\sigma^2$.

Based on the efficient formulas for calculating the $PRESS(\lambda)$ function, both these model selection alternatives can be implemented without affecting the total computational costs significantly.

\begin{figure}
\centering
\includegraphics[scale=0.70]{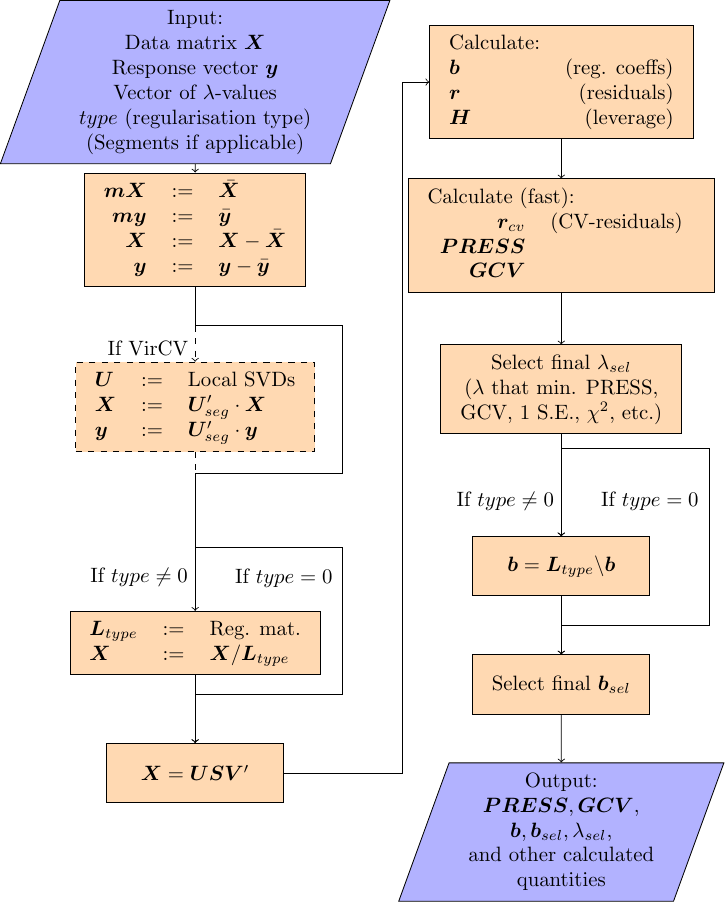}
\caption{{\color{black}Flow chart illustrating the LooCV, segmented CV, and VirCV. Most of the steps are common to all the algorithms. For the virtual CV we calculate local SVDs for each segment and left multiply by the transposed left-singular vectors of the segments prior to applying the regularisation matrix (if any). Detailed calculations and minor differences between the algorithms such as the modified leverage correction for VirCV are not shown.}}
\label{fig:tikz-greier}
\end{figure}

\section{Applications}\label{Sec: applications}
In the following, we demonstrate some applications of our fast cross-validation approaches for model selection within the TR framework for several real-world datasets. 
We consider situations where both leave-one-out and segmented cross-validation are appropriate. 
The required algorithms were implemented and executed in MATLAB, and prototype code is given in Appendices \ref{trcode}-\ref{VirCVcode}. A corresponding implementation in R-code will be made available upon publication at https://CRAN.R-project.org/package=TR. We used a computer running Mac OS Ventura 13.0.1 and MATLAB R2022a, with 16\,GB RAM, and an M1 Pro 10-core processor. 
For the derivative regularisation, we use the full rank approximations described in Section \ref{secL2} with the scaling coefficient set to $\epsilon=10^{-10}$ in the appended rows in the discrete regularisation matrices. This is done to mitigate the numerical impact from these rows in the resulting regression coefficients.

\subsection{The fast leave-one-out cross-validation}
\subsubsection{Datasets}
The following datasets will be considered in the examples presented below:

\begin{enumerate}
\item \textit{Octane data} \cite{kalivas97}. This dataset consists of near-infrared (NIR) spectra of gasoline. There are $60$ samples and $401$ features (wavelengths in the range $900\,nm-1700\,nm$). The response value is the octane number measured for each sample.
\item \textit{Pork fat data} \cite{lyndgaard12}. This dataset consists of Raman spectra measured on pork fat tissue. There are $105$ samples, $5567$ features (wavenumbers in the range $1889.9\,cm^{-1}-200.1\,cm^{-1}$), and $19$ different responses. For modelling and prediction, we only consider the response consisting of saturated fatty acids as a percentage of total fatty acids, hereafter referred to as SFA.
\item \textit{Prostate gene data} \cite{singh02}. The dataset is a microarray gene expression dataset. There are $102$ samples, and the gene expression of $12600$ different genes were measured. The response is binary (cancer/not cancer), and we consider the dummy-regression approach to the underlying classification problem. For this dataset, we standardise the data prior to modelling. The standardisation will introduce a small bias in the model selection that will be discussed later. 
\end{enumerate}
\noindent
For all datasets, we have used approximately $2/3$ of the available samples for model building and -selection. The remaining $1/3$ of the samples were used for testing the selected models. {(\color{black}Note that our choice of data splitting is somewhat arbitrary, just to serve the purpose of illustrating the ideas with an appropriate number of samples for both training and testing.)} We considered the following model selection alternatives identifying good regularisation parameter candidates: (i) $PRESS_{min}$ -- the minimum $PRESS(\lambda)$-value, (ii) $GCV_{min}$ -- the minimum $GCV(\lambda)$-value, (iii) the $1$ standard error rule for $PRESS(\lambda)$, (iv) the $\chi^2$-rule for $PRESS(\lambda)$ using the significance level $\alpha=0.2$.

\subsubsection{Model selection and prediction}
\begin{figure}[!htb] 
\centering 
 \includegraphics[width=0.50\textwidth]{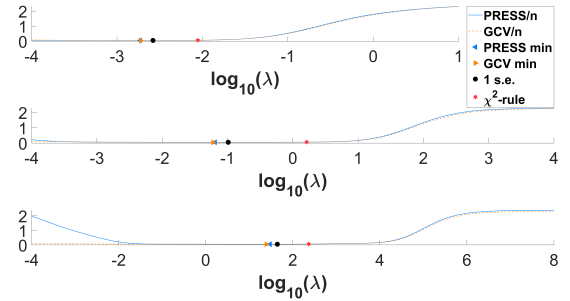} 
\caption{\textit{Octane data}. $PRESS/n$ and $GCV/n$ for a range of regularisation parameter values and different regularisation matrices. \textit{Top}: $L_2$ regularisation. \textit{Middle}: 1st derivative regularisation. \textit{Bottom}: 2nd derivative regularisation. The minimum $PRESS$ and $GCV$ values have been marked, as well as the regularisation parameter values selected by the $1$ S.E. rule and the $\chi^2$-rule.}
\label{fig:octanermsecv}
\end{figure}

\begin{figure}[!htb] 
\centering 
\includegraphics[width=0.5\textwidth]{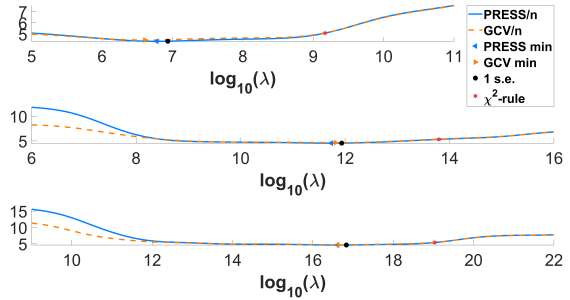} 
\caption{\textit{Pork fat data and SFA response.} $PRESS/n$ and $GCV/n$ for a range of regularisation parameter values and different regularisation matrices. \textit{Top}: $L_2$ regularisation. \textit{Middle}: 1st derivative regularisation. \textit{Bottom}: 2nd derivative regularisation. The minimum $PRESS$ and $GCV$ values have been marked, as well as the regularisation parameter values selected by the $1$ S.E. rule and the $\chi^2$-rule.}
\label{fig:porkrmsecv}
\end{figure}

\begin{figure}[!htb] 
\centering 
\includegraphics[height=0.22\textwidth]{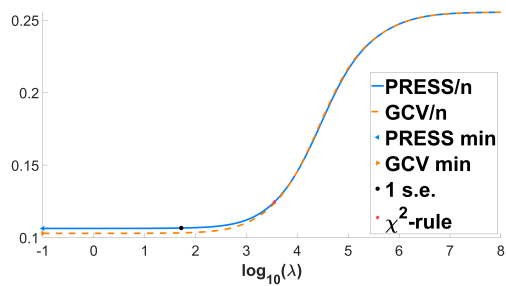} 
\caption{\textit{Prostate data.} $PRESS/n$ and $GCV/n$ for a range of regularisation parameter values using $L_2$ regularisation. The minimum $PRESS$ and $GCV$ values have been marked, as well as the regularisation parameter values selected by the $1$ S.E. rule and the $\chi^2$-rule.}
\label{fig:prostatermsecv}
\end{figure}

For each dataset, the modelling was based on {\color{black}a grid search of} $1000$ regularisation parameter candidate values spaced uniformly on a log-scale. For the octane data, the displayed values were in the range $10^{-4}$ to $10^5$, for the Pork fat data in the range $10^{2}$ to  $10^{25}$, and for the Prostate data in the range $10^{-1}$ to $10^{8}$. Different ranges were chosen for each dataset to avoid 
irrelevant levels of regularisation, and to obtain a good visualisation of the $PRESS$- and $GCV$ curves including the located minima.
In Figures \ref{fig:octanermsecv}--\ref{fig:prostatermsecv} the $PRESS/n$ and $GCV/n$ are plotted as functions of the regularisation parameter for the different datasets and the different choices of the regularisation matrix. Such plots are useful for model selection as they allow for a direct comparison of the model quality for different values of the regularisation parameter. 
Division of the $PRESS$- and $GCV$ values by the sample size $n$ makes the model selection statistics directly comparable to the prediction results obtained by the test sets. The test set results are shown in the Tables \ref{tab:rmsepoctane}--\ref{tab:rmsepprostate}. 

{\color{black}For the prostate data, the percentage correctly classified on the training set using cross-validation (classifying each sample to the largest of the fitted target values when using $0/1$ dummy-coding for the group memberships) is $91.2\%$ for all the parameter selection methods (it should be noted that this number happens to be identical to the test set result for most of the parameter selection methods).}

It should be noted that most of the displayed $PRESS$- (and $GCV$-) curves are relatively flat without a very distinct minimum point. Therefore it may be advantageous to employ either the $1$ S.E. rule or the $\chi^2$-rule to assure the selection of a  simpler model. For the Prostate data, in particular, we note that the smallest available candidate regularisation parameter value provides the minimum $PRESS$-value. The effect in terms of prediction when using the $1$ S.E. rule or the $\chi^2$-rule to obtain a simpler model varies between the datasets. For the Pork fat data, the $\chi^2$-rule gives better prediction than the other parameter selection methods for the SFA response, while the $\chi^2$-rule selects a poorer model than the other parameter selection methods on the Prostate data.

For the most precise identification of the $PRESS$- and $GCV$-minima a numerical optimiser should be used. However, in most practical situations the suggested strategy of considering just a subset of candidate regularisation parameter values is usually good enough for approximating the minima before doing the subsequent identification of parsimonious models (based on the principle of Occam's razor) that predict well.




\begin{table*}
\centering 
\begin{tabular}{|c|c|c|c|} \hline
\backslashbox{Parameter selection method}{Regularisation type} & $L_2$ & First derivative & Second derivative \\ \hline
Minimum PRESS value & 0.057  & 0.047 & 0.038 \\ \hline
Minimum GCV value &  0.057  & 0.047 & 0.039 \\ \hline
PRESS and 1 standard error rule & 0.059 & 0.045 & 0.036\\ \hline
PRESS and $\chi^2$-rule & 0.073 & 0.047 & 0.039 \\ \hline
\end{tabular}
\caption{\textit{Octane data.} $MSE$ (from test data) using various regularisation types and parameter selection methods.}
\label{tab:rmsepoctane}
\end{table*}

\begin{table*}
\centering 
\begin{tabular}{|c|c|c|c|} \hline
\backslashbox{Parameter selection method}{Regularisation type} & $L_2$ & First derivative & Second derivative \\ \hline
Minimum PRESS value & 4.46 & 5.39 & 5.56 \\ \hline
Minimum GCV value & 4.36   & 5.45 & 5.58 \\ \hline
PRESS and 1 standard error rule &  4.58  & 5.56  & 5.72 \\ \hline
PRESS and $\chi^2$-rule &  4.11  & 4.32 & 4.20\\ \hline
\end{tabular}
\caption{\textit{Pork fat data.} $MSE$ (from test data) for the SFA response using various regularisation types and parameter selection methods.}
\label{tab:rmsepporkfatSFA}
\end{table*}



\begin{table}
\centering 
\begin{tabular}{|c|c|} \hline
Parameter selection method & PCC test set\\ \hline
Minimum PRESS value & 91.2 \\ \hline
Minimum GCV value & 91.2  \\ \hline
PRESS and 1 standard error rule & 91.2 \\ \hline
PRESS and $\chi^2$-rule &  88.2 \\ \hline
\end{tabular}
\caption{\textit{Prostate data.} Percentage of correctly classified (PCC) 
samples using the test set predictions of the selected $0 - 1$ dummy regression model based on $L_2$ regularisation.} 
\label{tab:rmsepprostate}
\end{table}

\subsubsection{Regression coefficients}
\begin{figure}[!htb] 
\centering 
\includegraphics[height=0.25\textwidth]{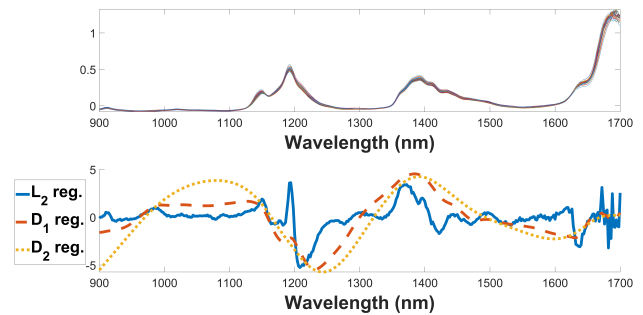} 
\caption{\textit{Octane data. Top}: Plot of the NIR spectra of octane. \textit{Bottom}: $PRESS$-minimal regression coefficients based on different regularisation matrices.}
\label{fig:octanespec}
\end{figure}

Figure \ref{fig:octanespec} shows the octane data together with the $PRESS$-minimal regression coefficients using the $L_2$-, the first derivative-, and the second derivative regularisations. 
Note that the choice of regularisation matrix heavily influences the appearance of the regression coefficients without the minimum $PRESS$- or $GCV$ values changing much.
Table \ref{tab:rmsepoctane} confirms that the predictive powers are relatively similar for all these models.  Doing consistent model interpretations solely based on the regression coefficients in Figure \ref{fig:octanespec} is obviously a challenging (if not impossible) task, see also \cite{brown09}.

\subsubsection{Computational speed}
Table \ref{tab:time} shows the computation times for model selection with the different datasets and different types of regularisation when varying the number of regularisation parameter candidate values {\color{black}(varying the number of points in the search grid)}. 
The times in Table \ref{tab:time} also include the computation of the regression coefficients corresponding to the minimal $GCV$ and $PRESS$ values for all responses.
The main differences in computational time between finding the SVD in the case of $L_2$ regularisation and in the cases of first- and second-derivative regularisation are due to the initial calculations of
$\tilde{\bX}$, see Section \ref{secL2}. Similarly, the required transformation of the regression coefficients (see \eqref{backtransf}) explains the increase in computational time from calculating the SVD only to finding $PRESS$, $GCV$ and regression coefficients for a single regularisation parameter value for the first and second derivative regularisation. 

\begin{table*}[!htb]
\centering 
\begin{tabular}{|c|c|c|c|c|c|c|} \hline
\backslashbox{Data (reg. type)}{Number of $\lambda$-values} & $0$ (SVD only) & $1$ & $10$ & $100$ & $1000$ & $10000$ \\ \hline
Octane ($L_2$) & 0.0014 & 0.0014 & 0.0014 & 0.0016 & 0.0024 & 0.013 \\ \hline
Octane (1st derivative) & 0.0034 & 0.0046 & 0.0051 & 0.0052 & 0.0055 & 0.017 \\ \hline
Octane (2nd derivative) & 0.0048 & 0.0074 & 0.0082 & 0.0082 & 0.0087 & 0.020 \\ \hline
Pork fat ($L_2$) & 0.018 & 0.023 & 0.023 & 0.026 & 0.040 & 0.26 \\ \hline
Pork fat (1st derivative) & 0.096 & 0.22 & 0.22 & 0.22 & 0.24 & 0.46 \\ \hline
Pork fat (2nd derivative) & 0.23 & 0.59 & 0.60 & 0.62 & 0.64 & 0.85 \\ \hline
Prostate ($L_2$) & 0.038 & 0.072 & 0.077 & 0.078 & 0.078 & 0.11 \\ \hline
\end{tabular}
\caption{\textit{Computing time} (in seconds) for model selection including finding the $PRESS$- and $GCV$-minimal regression coefficients when varying the number of candidate regularisation parameter values. The times are the averages of $50$ repeated runs rounded to the two most significant digits.}
\label{tab:time}
\end{table*}

\subsection{Segmented cross-validation}

\subsubsection{Datasets}
\begin{figure*}[!htb] 
\centering 
\includegraphics[width=0.95\textwidth]{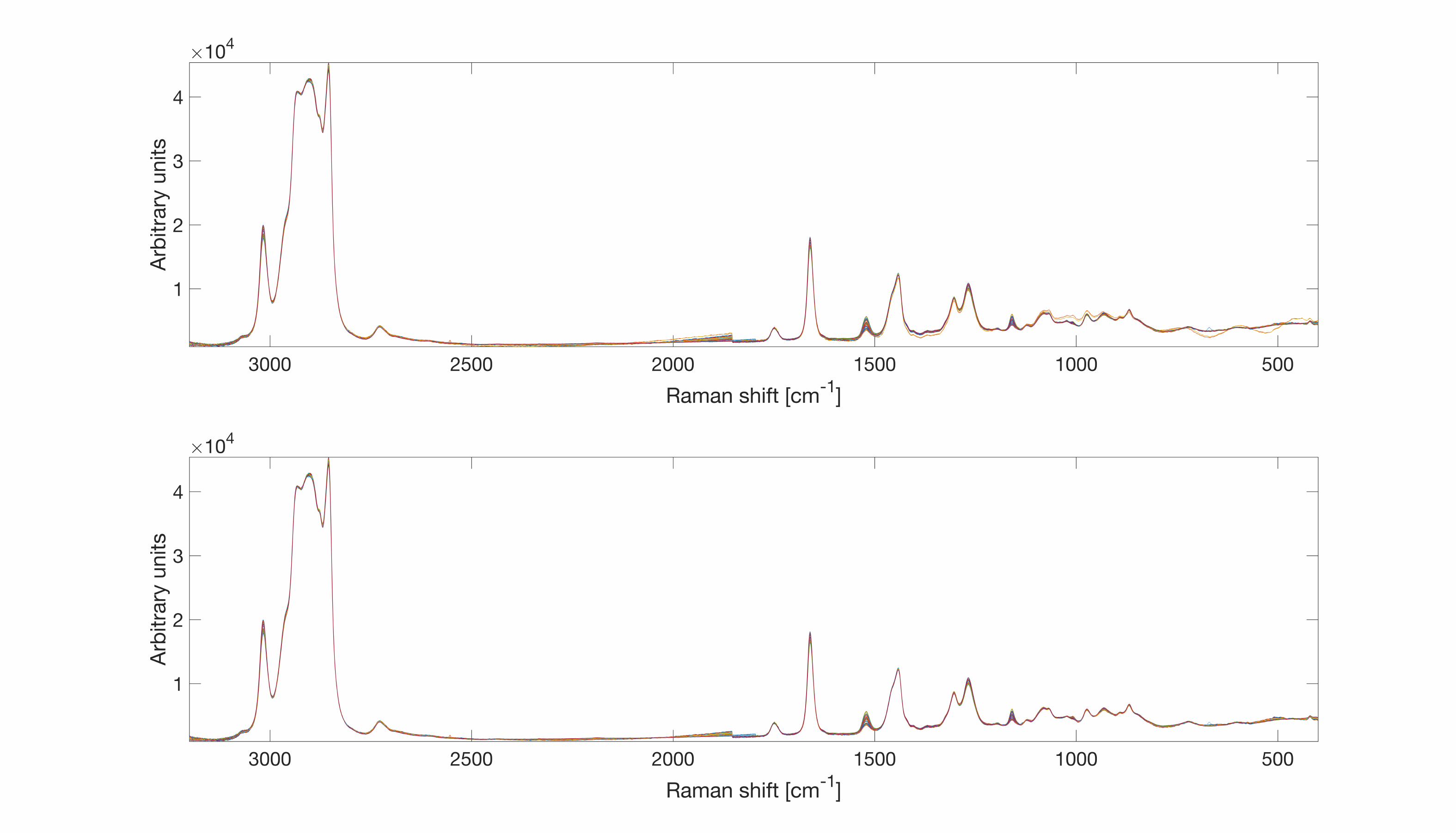} 
\caption{\textit{Plot of the fish oil spectra} after pre-processing with EMSC with 6th order polynomial baseline (\textit{top}) and additional replicate correction (\textit{bottom}).}
\label{fig:fishspec}
\end{figure*}

\begin{figure}[!htb] 
\centering 
\includegraphics[width=0.5\textwidth]{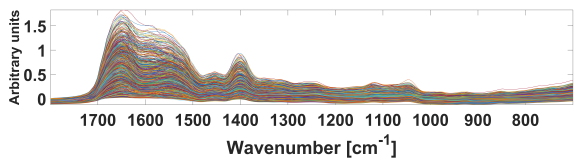} 
\caption{\textit{Plot of the hydrolysis spectra} after pre-processing with EMSC with 2nd order polynomial baseline.}
\label{fig:hydrolysisspec}
\end{figure}

\begin{figure}[!htb] 
\centering 
\includegraphics[width=0.5\textwidth]{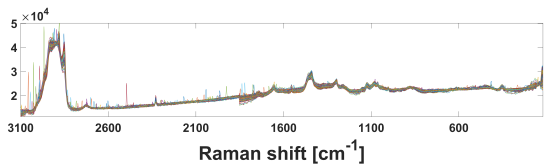} 
\caption{\textit{Plot of the milk spectra} after pre-processing with EMSC with 6th order polynomial baseline. Noise in some replicates is clearly visible as spikes around the main variation. }
\label{fig:milkspec}
\end{figure}

In the following we will demonstrate the use of segmented cross-validation with $L_2$ regularisation for three datasets:

\begin{enumerate}
\item Raman spectra of fish oil \cite{afseth06}. The dataset consists of $42$ sample segments including $3$ replicate spectra of each unique sample giving a total of $126$ rows and $2801$ wavenumbers in the range $3200\,cm^{-1}$ to $400\,cm^{-1}$. The response variable was the iodine value (the response values were identical across each segment), which is frequently used as an indicator of the degree of unsaturation of fat \cite{afseth06}. The spectra of this dataset are plotted in Figure \ref{fig:fishspec} after applying Extended Multiplicative Signal Correction (EMSC) \cite{afseth12} with 6th-order polynomial baseline correction.
\item {\color{black}Fourier transform infrared (FTIR) spectra of hydrolysates from various mixtures of rest raw materials and enzymes \cite{kristoffersen2019ftir}. The dataset consists of $332$ samples including 1 to 12 replicates of each unique sample giving a total of $885$ rows and $571$ wavenumbers in the range $1800\,cm^{-1}$ to $700\,cm^{-1}$. The response variable was average molecular weight (AMW) (identical across each replicate set), which can be used as a proxy for the degree of hydrolysation. The spectra of this dataset are plotted in Figure \ref{fig:hydrolysisspec}.}
\item Raman milk spectra \cite{afseth10,randby12,liland16}. The dataset consists of $232$ unique sample segments including between $6$ and $12$ replicate measurements of each unique sample giving a total of $2682$ rows and $2981$ wavenumbers in the range $3100\,cm^{-1}$ to $120\,cm^{-1}$. The response variables were the iodine value and the concentration of conjugated linoleic acid (CLA). Also for this dataset, the response values were identical across each segment. The spectra of this dataset are plotted in Figure \ref{fig:milkspec} after applying EMSC with 6th-order polynomial baseline correction.

\end{enumerate}
\noindent
For all datasets, we have excluded the endpoint regions of the original spectra due to noise and the poor quality of the measurements. The wave numbers reported above are those included after this truncation.
Approximately 2/3 of the replicate segments were used for model building and -selection, and the remaining 1/3 of the segments were used as a test set. 

The following four model selection strategies were considered: (i) $PRESS_{min}$ -- the minimum $PRESS(\lambda)$-value from LooCV (ignoring the presence of sample segments), (ii) $GCV_{min}$ -- the minimum $GCV(\lambda)$-value, (iii) the $PRESS_{min}$ from the SegCV (successively holding out the entire sample segments), and (iv) the $PRESS_{min}$ from the VirCV. 
We have chosen to focus only on the parameter selections associated with the minima of the various error curves in this part of our study
(neither the $\chi^2$-rule nor the $1$ S.E. rule turned out to affect the model selections much). {\color{black}Neither of the two strategies for quicker estimation of $PRESS$-values is shown in the plots as the minimum $PRESS$-value (from searching) coincides with the minimum-$PRESS$ value from the 1000 sampled $\lambda$-values and the cubic spline interpolation is visually indistinguishable from the full $PRESS$ curve obtained from explicit segment removal.}

\subsubsection{Fish data -- effect of pre-processing}
\label{sec:fish}

Spectroscopic measurements may be corrupted by both additive and multiplicative types of noise. Pre-processing of such data prior to modelling is therefore usually required. It is therefore of particular interest also to investigate how the model selection strategies considered above compare for pre-processed data. 
In particular, we will consider the Extended Multiplicative Signal Correction (EMSC) \cite{afseth12} with replicate corrections \cite{kohler09}. 

In general, the goal of the EMSC pre-processing is to adjust all the measured spectra to a common scale and to eliminate the possible effects of additive noise. This includes the estimation of an individual scaling constant for each spectrum and an orthogonalisation step that de-trends the spectra with respect to some set of lower-order polynomial trends (the reader is referred to the provided references for the technical details). In the present examples with Raman spectra, the samples were orthogonalised with respect to the subspace including all polynomial trends up to the $6$-th degree. 

The Raman spectra of fish samples were subjected to EMSC pre-processing to compensate for different scaling and competing phenomena such as fluorescence and optical/scattering effects in the equipment and samples. For the milk data, the spectrum having the least fluorescence background was chosen as a reference, though the effect of choice of reference spectrum is minimal.

For datasets including segments of replicated measurements, a replicate correction step is often considered to alleviate the presence of inter-replicate variance. Such correction can be done by 
an initial EMSC-based pre-processing of the spectra in each sample segment. Thereafter, the corrected sample segments can be individually mean-centred and organised into a full data matrix.

As we expect the dominant right singular vectors of the full matrix to account for the most dominant inter-replicate variance, orthogonalisation of the data with respect to one or more of the associated dimensions contributes to making the replicates more similar, see \cite{kohler09} for details. Because every sample in the training dataset is included in the pre-processing, some bias affecting the subsequent $PRESS$-calculations and model selection must be expected.

\begin{figure}[!htb]
\centering 
\includegraphics[height=0.25\textwidth]{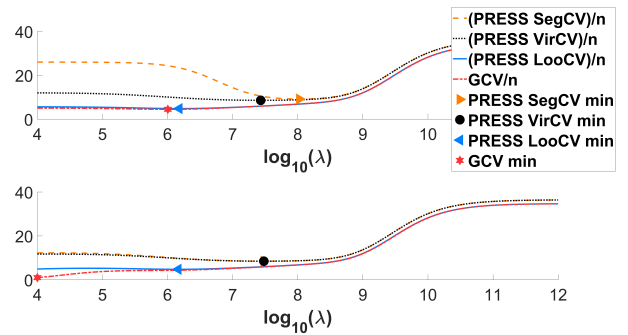} 
\caption{\textit{Fish oil data.} Model selection for data pre-processed with the EMSC both with and without replicate correction. Top: Standard EMSC pre-processing. Bottom: EMSC with 30\% of the inter-replicate variance removed.}
\label{fig:fishrep}
\end{figure}

Figure \ref{fig:fishrep} shows the model selection for pre-processed fish oil data based on the pure EMSC and for the EMSC where $30\%$ of the inter-replicate variance is removed. It is evident that the SegCV and the VirCV become considerably more similar in the latter case.
As one should expect, the $GCV$- and $PRESS$ curves based on the LooCV seem to provide unrealistically low error values and the selection of lesser regularised models.
This phenomenon does not occur with the SegCV 
where an entire segment of replicates is held out in each cross-validation step. The VirCV seems quite robust against the inter-replicate variance.


\begin{table*}[!htb]
\centering 
\begin{tabular}{|c|c|c|c|c|} \hline
\backslashbox{Pre proc.}{Selection curves} & LooCV & GCV & VirCV & SegCV\\ \hline
Raw data & 20.3 & 21.5 & 12.3 & 9.7\\ \hline
EMSC & 14.4 & 15.1 & 6.9 & 4.5 \\ \hline
EMSC + 30\% inter-replicate variance removed & 14.4 & 15.9 & 6.7 & 6.7\\ \hline
\end{tabular}
\caption{\textit{Fish oil data. $MSE$ (from test data) for different model selection strategies and different pre-processing alternatives.}}
\label{tab:fishoilrmsep}
\end{table*}
\noindent
The prediction results for the test set of the fish oil data with the various pre-processing alternatives are presented in Table \ref{tab:fishoilrmsep},
and shows that the best results are obtained with the ordinary EMSC pre-processing and model selection based on the SegCV.
By simultaneously considering Figure \ref{fig:fishrep}, it is clear that the more heavily regularised among the selected models (those based on the largest regularisation parameter values) perform better on the test set. With standard EMSC pre-processing the minima of the VirCV is located at a smaller regularisation parameter value than for the SegCV, suggesting an explanation of the difference in predictive performance.

For the milk data, the prediction error estimates obtained after pre-processing the data are similar for all the parameter selection methods (table omitted), as was also the case with the raw data. 

\subsubsection{Hydrolysis data -- heterogeneous segments}
\label{sec:hydrolysis}
{\color{black}The hydrolysis data is used as an example of a model comparison which is often performed using 5-fold or 10-fold segmented cross-validation. For the FTIR data, we have chosen a 5-fold strategy where replicates are kept together inside each fold to prevent information bleeding by replicates of the same sample appearing in both training and test data. The resulting cross-validation segments vary in size from 103 to 117 samples, each, due to the present replicate sets. We have chosen to combine this with a 2nd derivative regularisation.}

\begin{figure}[!htb] 
\centering 
\includegraphics[height=0.13\textwidth]{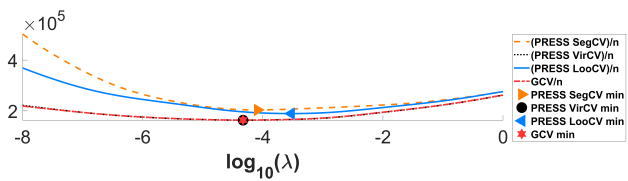} 
\caption{\textit{Hydrolysis data.} Different model selection strategies for a range of regularisation parameter values using 2nd derivative regularisation.}
\label{fig:hydrolysisrmsecv}
\end{figure}

In Figure \ref{fig:hydrolysisrmsecv}, we have plotted the $PRESS$-curves for SegCV, VirCV, LooCV and GCV. For these highly heterogeneous cross-validation segments, the virtual cross-validation strategy coincides with $GCV$, both underestimating the prediction errors. Also, LooCV underestimates the errors, but less so. Since the general forms of the $PRESS$-curves are quite similar, the minimum $PRESS$-values are located quite close together, suggesting that for the FTIR dataset, any of the strategies will give a reasonable estimate of the optimal $\lambda$-value. As Table \ref{tab:amwrmsep} suggests, performance when applying the regressions corresponding to minimal $PRESS$-values on the test data are also similar with a slight advantage to the more regularised LooCV solution.

\begin{table}[!htb]
\centering 
\begin{tabular}{|c|c|c|c|c|} \hline
\backslashbox{Pre proc.}{Selection curves} & LooCV & GCV & VirCV & SegCV\\ \hline
EMSC & 1.85 & 1.92 & 1.92 & 1.89 \\ \hline
\end{tabular}
\caption{\textit{Hydrolysis data. $MSE$ (from test data) using EMSC for pre-processing.}}
\label{tab:amwrmsep}
\end{table}

\subsubsection{Milk data -- efficiency with many segments}
\label{sec:milk}
\begin{figure}[!htb] 
\centering 
\includegraphics[height=0.25\textwidth]{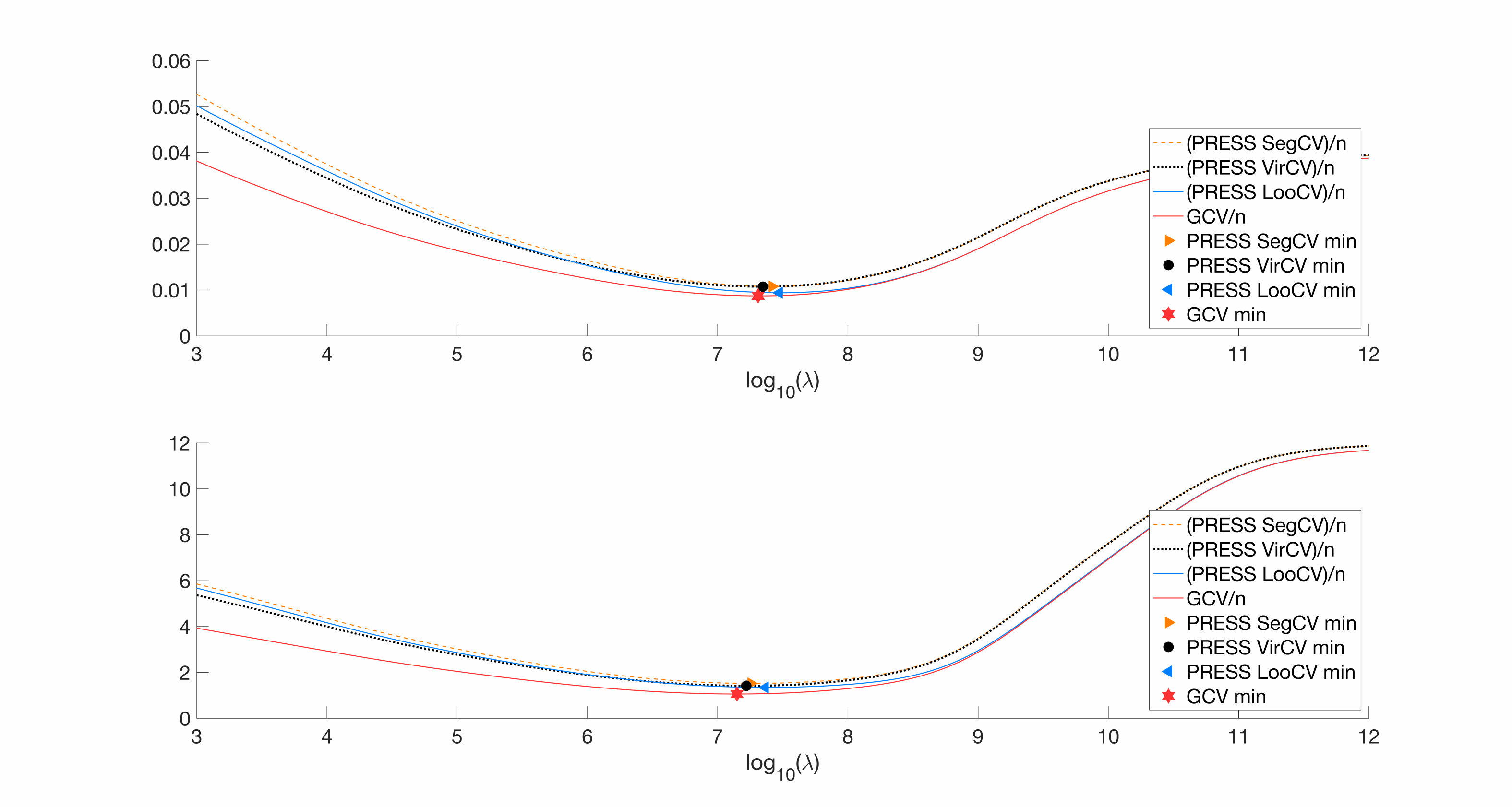} 
\caption{\textit{Milk data.} Different model selection strategies for a range of regularisation parameter values using $L_2$ regularisation. \textit{Top}: CLA. \textit{Bottom}: Iodine value.}
\label{fig:milkrmsecv}
\end{figure}

The milk data is an example of relatively many samples (2682) and replicate groups (232), which can be challenging with regard to computational resources when cross-validating over a large range of $\lambda$-values. As can be observed from Figure \ref{fig:milkrmsecv}, the differences between SegCV, VirCV, LooCV and GCV are small both with regard to the shape of the curves and location of respective minimum values. This is due to the low variation between samples within each replicate group, in sharp contrast to the FTIR dataset with its highly heterogeneous cross-validation segments. Of more interest, is the time usage for the various strategies, which is summarised in Section \ref{sec:speed} below.

\subsubsection{Approximations of $PRESS$-values - computational speed}\label{sec:speed}
Table \ref{tab:timeblockorth} shows the computational times for the different model selection strategies. Both the $PRESS$- and the $GCV$ values are included as computing only one of them takes approximately the same time as computing both. Because the size of the replicate segments are relatively small for the Raman datasets ($3$ replicate measurements for the fish oil data and $6$ to $12$ replicate measurements for the milk data), the SVDs required for the internal orthogonalisations of the segments contribute insignificantly to the total computational load. The amount of computations required for model selection based on the VirCV is therefore quite comparable to the computations required for the LooCV version of $PRESS$ (and for the $GCV$). {\color{black}The strategy of searching for the minimum $PRESS$-value by golden section search and parabolic interpolation (MinSearch), is remarkably similar to VirCV in time usage. However, there is a trade-off between obtaining an estimate of the exact minimum value (MinSearch) and a full $PRESS$-curve (VirCV). Approximation of the SegCV using spline interpolation is slower than VirCV and MinSearch, but still sufficiently fast for practical use in all tested cases and with the advantage of giving a $PRESS$-curve highly similar to the one obtained by the SegCV. The implicit segmented cross-validation (ImpCV) using Theorem \ref{TmSegCV}, is faster than SegCV for small segments and a bit slower for large segments, though still fast enough to provide exact results for all $\lambda$ values. In general, the initial calculation of the SVD seems to be the main limiting factor in computational speed when the datasets grow in size. This is especially prominent for the milk data where SegCV performs this initial SVD 232 times. Here, a strategy avoiding SVD or using a randomised SVD algorithm \cite{Halko2011} might be favourable, however, the other presented strategies are still usable.}

\begin{table*}[!htb]
\centering 
\begin{tabular}{|c|c|c|c|c|c|c|c|} \hline
Dataset & SegCV & ImpCV & VirCV & MinSearch & Spline & PRESS\&GCV  \\ \hline
Fish oil & $0.624 ~(1)$ & $0.100 ~(1/6)$ & $0.016 ~(1/38)$ & $0.015 ~(1/42)$ & $0.062 ~(1/10)$ & $0.010 ~(1/65)$\\ \hline 
Hydrolysis & $0.686 ~(1)$ & $0.927 ~(1/0.7)$ & $0.106 ~(1/6)$ & $0.118 ~(1/6)$ & $0.252 ~(1/3)$ & $0.096 ~(1/7)$\\ \hline 
Milk & $867.8 ~(1)$ & $8.9 ~(1/98)$ & $3.3 ~(1/266)$ & $3.2 ~(1/271)$ & $4.1 ~(1/214)$  & $2.95 ~(1/294)$ \\ \hline 
\end{tabular}
\caption{\textit{Computational time for different model selection strategies for the Fish oil data, Hydrolysis data and Milk data when considering $500$ candidate regularisation parameter values. The times are given in seconds, rounded to two significant digits, and is the average of $50$ repeated runs. The speedup relative to SegCV is shown in parenthesis.}}
\label{tab:timeblockorth}
\end{table*}

\section{Discussion and conclusions}
The essence of the TR-framework described in the present work is that just a single SVD-calculation (of either the original data matrix $\bX$ or a transformed version $\tilde{\bX}$) is required to explore some particular regularised regression problem of interest.
We have pointed out that the $PRESS$- and $GCV$ values required for model selection(s) based on the LooCV or the $GCV$ can be obtained at the computational cost of two matrix-vector multiplications for each choice of the regularisation parameter value $\lambda$. In the applications section, it is demonstrated that our framework scales well when increasing the number of candidate regularisation parameter values
in the case of 'small $n$ with large $p$' problems. {\color{black}This scaling will also work well for problems involving multiple responses as most of the computations will be shared among responses.} For smaller and medium-sized data as well as for other situations where the required SVD can be calculated (or approximated) reasonably fast, the acquired computational efficiency allows for the exploration of a large number of candidate models in a very short amount of time.

{\color{black}
For situations where leave-one-out cross-validation underestimates validation error because of sample replicates or another grouping of samples, segmented cross-validation is the appropriate choice. We have proved a theorem saying that explicit remodelling for computation of cross-validated $PRESS$-values can be avoided, while still giving exact results, at the computational cost of inverting one matrix per sample segment per $\lambda$-value. For cases where the cost outweighs the benefits, we have proposed alternative strategies for reducing the number of inversions through careful selections of $\lambda$-values as well as an approximate virtual cross-validation (VirCV) strategy. T}he VirCV is a computationally efficient approximation of the traditional SegCV. In the applications (Section \ref{Sec: applications}) we observed that the VirCV approximation of the SegCV appears to be quite accurate for model selection in the case of highly similar samples within each segment while using the LooCV or $GCV$ in such situations is more likely to propose insufficient regularisation and models that predict poorer.

It is important to note that when the dataset is pre-processed and/or transformed by a data-dependent method, some bias both in the LooCV- and VirCV-based $PRESS$ values must be expected. The data variable standardisation commonly used in RR is a typical example. The EMSC pre-processing that was used with or without replicate corrections is another. However, the main purpose of the LooCV- and VirCV-based $PRESS$ values in the proposed framework is model selection rather than error estimation. The bias introduced by such pre-processing methods is therefore not likely to be very harmful as long as the (training) data does not contain serious outliers.

Although leverage correction of the model residuals for fast calculation of the LooCV in linear least squares regression problems is well known, there are some misleading assertions in the literature regarding both the properties and accuracy of $PRESS$-values that require clarification: 
i) Hansen\cite[page 96]{hansen10} claims that the leverage values are not invariant under row permutations of the $\bX$-data making the $PRESS$-values dependent on the ordering of the data.
However, when the rows of the data matrix are permuted it can be verified that the leverage values are unchanged and undergo precisely the same permutation. Consequently, the correct leverage values will match up perfectly with the corresponding model residuals in the calculation of the $PRESS(\lambda)$ calculations assuring its invariance under any row permutation of the $(\bX,\by)$-data. 
ii) Myers\cite[page 399]{myers90} claims that the expression for fast calculation of $PRESS(\lambda)$ is only an approximation when performing centring and scaling of the data. This is, however, only true when the scaling factors are calculated from the data to be used in the model building. The data centring, as such, does not corrupt the leverage- and $PRESS(\lambda)$-values as long as the $1/n$ terms are included in the associated leverage corrections of the model residuals. 
iii) The version of Ridge regression implemented in the MASS package\cite{MASS} for the R programming language includes a fast calculation of the $GCV(\lambda)$-values for a desired vector of corresponding $\lambda$-values. The $1/n$ term is, however, ignored when correcting the model residuals by the required averaged leverage value. Consequently, the resulting $GCV$-values are misleading when the centring of the data is included as a part of the Ridge Regression modelling.

We believe that future statistical texts and software dealing with Ridge Regression (and Tikhonov Regularisation) will find value in including the necessary pieces of linear algebra (in particular the simple matrix-vector multiplications of Equation (\ref{hlambda}) to establish the fast calculation of the $PRESS(\lambda)$ in Equation (\ref{PRESSlambda}). In our opinion, these relatively simple but still powerful results demonstrate yet another remarkable consequence of the SVD at the core of applied multivariate data analysis.

{\color{black} Finally, we have established a theorem describing how to compute the cross-validated residuals for (regularised) linear regression models from the fitted value residuals. The computation can be seen as a multi-sample kind of leverage correction that applies to any type of segmented cross-validation strategy. In many cases, it represents a computationally efficient alternative to the computationally slower "hold out/remodelling approach" most common within statistics and machine learning. For the special case of LooCV, our theorem simplifies to the well-known scalar leverage correction calculations of the LooCV errors.}

\bibliographystyle{IEEEtran}  
\bibliography{Bibliography.bib}  

\begin{thebibliography}{10}
\providecommand{\url}[1]{#1}
\csname url@samestyle\endcsname
\providecommand{\newblock}{\relax}
\providecommand{\bibinfo}[2]{#2}
\providecommand{\BIBentrySTDinterwordspacing}{\spaceskip=0pt\relax}
\providecommand{\BIBentryALTinterwordstretchfactor}{4}
\providecommand{\BIBentryALTinterwordspacing}{\spaceskip=\fontdimen2\font plus
\BIBentryALTinterwordstretchfactor\fontdimen3\font minus
  \fontdimen4\font\relax}
\providecommand{\BIBforeignlanguage}[2]{{%
\expandafter\ifx\csname l@#1\endcsname\relax
\typeout{** WARNING: IEEEtran.bst: No hyphenation pattern has been}%
\typeout{** loaded for the language `#1'. Using the pattern for}%
\typeout{** the default language instead.}%
\else
\language=\csname l@#1\endcsname
\fi
#2}}
\providecommand{\BIBdecl}{\relax}
\BIBdecl

\bibitem{hastie09}
J.~Friedman, T.~Hastie, and R.~Tibshirani, \emph{The elements of statistical
  learning}.\hskip 1em plus 0.5em minus 0.4em\relax Springer series in
  statistics Springer, Berlin, 2009, vol.~1.

\bibitem{hjort93}
J.~S.~U. Hjorth, \emph{Computer Intensive Statistical Methods: Validation,
  Model Selection and Bootstrap}.\hskip 1em plus 0.5em minus 0.4em\relax
  Chapman and Hall/CRC, 1993.

\bibitem{stone74}
M.~Stone, ``Cross-validatory choice and assessment of statistical
  predictions,'' \emph{Journal of the royal statistical society. Series B
  (Methodological)}, pp. 111--147, 1974.

\bibitem{golub79}
G.~H. Golub, M.~Heath, and G.~Wahba, ``Generalized cross-validation as a method
  for choosing a good ridge parameter,'' \emph{Technometrics}, vol.~21, no.~2,
  pp. 215--223, 1979.

\bibitem{benkeser2020improved}
D.~Benkeser, M.~Petersen, and M.~J. van~der Laan, ``Improved small-sample
  estimation of nonlinear cross-validated prediction metrics,'' \emph{Journal
  of the American Statistical Association}, vol. 115, no. 532, pp. 1917--1932,
  2020.

\bibitem{bates2023cross}
S.~Bates, T.~Hastie, and R.~Tibshirani, ``Cross-validation: what does it
  estimate and how well does it do it?'' \emph{Journal of the American
  Statistical Association}, pp. 1--12, 2023.

\bibitem{smith2014correcting}
G.~C. Smith, S.~R. Seaman, A.~M. Wood, P.~Royston, and I.~R. White,
  ``Correcting for optimistic prediction in small data sets,'' \emph{American
  journal of epidemiology}, vol. 180, no.~3, pp. 318--324, 2014.

\bibitem{celisse2016stability}
A.~Celisse and B.~Guedj, ``Stability revisited: new generalisation bounds for
  the leave-one-out,'' \emph{arXiv preprint arXiv:1608.06412}, 2016.

\bibitem{hoerl70}
A.~E. Hoerl and R.~W. Kennard, ``Ridge regression: Biased estimation for
  nonorthogonal problems,'' \emph{Technometrics}, vol.~12, no.~1, pp. 55--67,
  1970.

\bibitem{tikhonov63}
A.~N. Tikhonov, ``On the solution of ill-posed problems and the method of
  regularization,'' in \emph{Doklady Akademii Nauk}, vol. 151, number 3.\hskip
  1em plus 0.5em minus 0.4em\relax Russian Academy of Sciences, 1963, pp.
  501--504.

\bibitem{hansen10}
\BIBentryALTinterwordspacing
P.~C. Hansen, \emph{Discrete Inverse Problems}.\hskip 1em plus 0.5em minus
  0.4em\relax Society for Industrial and Applied Mathematics, 2010. [Online].
  Available: \url{http://epubs.siam.org/doi/abs/10.1137/1.9780898718836}
\BIBentrySTDinterwordspacing

\bibitem{allen71}
D.~M. Allen, ``Mean square error of prediction as a criterion for selecting
  variables,'' \emph{Technometrics}, vol.~13, no.~3, pp. 469--475, 1971.

\bibitem{allen74}
D.~M. Allen, ``The relationship between variable selection and data
  augmentation and a method for prediction,'' \emph{Technometrics}, vol.~16,
  no.~1, pp. 125--127, 1974.

\bibitem{householder65}
A.~S. Householder, \emph{The theory of matrices in numerical analysis}, ser.
  Blaisdell book in the pure and applied sciences.\hskip 1em plus 0.5em minus
  0.4em\relax Blaisdell Pub. Co., 1965.

\bibitem{kalivas12}
J.~H. Kalivas, ``Overview of two-norm (l2) and one-norm (l1) tikhonov
  regularization variants for full wavelength or sparse spectral multivariate
  calibration models or maintenance,'' \emph{Journal of Chemometrics}, vol.~26,
  no.~6, pp. 218--230, 2012.

\bibitem{phillips62}
D.~L. Phillips, ``A technique for the numerical solution of certain integral
  equations of the first kind,'' \emph{Journal of the ACM (JACM)}, vol.~9,
  no.~1, pp. 84--97, 1962.

\bibitem{bjorck16}
{\AA}.~Bj{\"o}rck, \emph{Numerical methods in matrix computations}.\hskip 1em
  plus 0.5em minus 0.4em\relax Springer, 2016.

\bibitem{brent1973algorithms}
R.~P. Brent, ``Algorithms for minimization without derivatives, chap. 4,''
  1973.

\bibitem{indahl05}
U.~Indahl, ``A twist to partial least squares regression,'' \emph{Journal of
  Chemometrics}, vol.~19, no.~1, pp. 32--44, 2005.

\bibitem{kalivas97}
\BIBentryALTinterwordspacing
J.~H. Kalivas, ``Two data sets of near infrared spectra,'' \emph{Chemometrics
  and Intelligent Laboratory Systems}, vol.~37, no.~2, pp. 255 -- 259, 1997.
  [Online]. Available:
  \url{http://www.sciencedirect.com/science/article/pii/S0169743997000385}
\BIBentrySTDinterwordspacing

\bibitem{lyndgaard12}
L.~B. Lyndgaard, K.~M. S{\o}rensen, F.~Berg, and S.~B. Engelsen, ``Depth
  profiling of porcine adipose tissue by raman spectroscopy,'' \emph{Journal of
  Raman Spectroscopy}, vol.~43, no.~4, pp. 482--489, 2012.

\bibitem{singh02}
D.~Singh, P.~G. Febbo, K.~Ross, D.~G. Jackson, J.~Manola, C.~Ladd, P.~Tamayo,
  A.~A. Renshaw, A.~V. D'Amico, J.~P. Richie \emph{et~al.}, ``Gene expression
  correlates of clinical prostate cancer behavior,'' \emph{Cancer cell},
  vol.~1, no.~2, pp. 203--209, 2002.

\bibitem{brown09}
\BIBentryALTinterwordspacing
C.~D. Brown and R.~L. Green, ``Critical factors limiting the interpretation of
  regression vectors in multivariate calibration,'' \emph{TrAC Trends in
  Analytical Chemistry}, vol.~28, no.~4, pp. 506 -- 514, 2009. [Online].
  Available:
  \url{http://www.sciencedirect.com/science/article/pii/S0165993609000363}
\BIBentrySTDinterwordspacing

\bibitem{afseth06}
N.~K. Afseth, J.~P. Wold, and V.~H. Segtnan, ``The potential of raman
  spectroscopy for characterisation of the fatty acid unsaturation of salmon,''
  \emph{Analytica chimica acta}, vol. 572, no.~1, pp. 85--92, 2006.

\bibitem{afseth12}
N.~K. Afseth and A.~Kohler, ``Extended multiplicative signal correction in
  vibrational spectroscopy, a tutorial,'' \emph{Chemometrics and Intelligent
  Laboratory Systems}, vol. 117, pp. 92--99, 2012.

\bibitem{kristoffersen2019ftir}
K.~A. Kristoffersen, K.~H. Liland, U.~B{\"o}cker, S.~G. Wubshet, D.~Lindberg,
  S.~J. Horn, and N.~K. Afseth, ``Ftir-based hierarchical modeling for
  prediction of average molecular weights of protein hydrolysates,''
  \emph{Talanta}, vol. 205, p. 120084, 2019.

\bibitem{afseth10}
N.~K. Afseth, H.~Martens, {\AA}.~Randby, L.~Gidskehaug, B.~Narum,
  K.~J{\o}rgensen, S.~Lien, and A.~Kohler, ``Predicting the fatty acid
  composition of milk: A comparison of two fourier transform infrared sampling
  techniques,'' \emph{Applied spectroscopy}, vol.~64, no.~7, pp. 700--707,
  2010.

\bibitem{randby12}
{\AA}.~Randby, M.~R. Weisbjerg, P.~N{\o}rgaard, and B.~Heringstad, ``Early
  lactation feed intake and milk yield responses of dairy cows offered grass
  silages harvested at early maturity stages,'' \emph{Journal of Dairy
  science}, vol.~95, no.~1, pp. 304--317, 2012.

\bibitem{liland16}
K.~H. Liland, A.~Kohler, and N.~K. Afseth, ``Model-based pre-processing in
  raman spectroscopy of biological samples,'' \emph{Journal of Raman
  Spectroscopy}, vol.~47, no.~6, pp. 643--650, 2016.

\bibitem{kohler09}
A.~Kohler, U.~B{\"o}cker, J.~Warringer, A.~Blomberg, S.~Omholt, E.~Stark, and
  H.~Martens, ``Reducing inter-replicate variation in fourier transform
  infrared spectroscopy by extended multiplicative signal correction,''
  \emph{Applied spectroscopy}, vol.~63, no.~3, pp. 296--305, 2009.

\bibitem{Halko2011}
N.~Halko, P.-G. Martinsson, and J.~A. Tropp, ``Finding structure with
  randomness: Probabilistic algorithms for constructing approximate matrix
  decompositions,'' \emph{SIAM review}, vol.~53, no.~2, pp. 217--288, 2011.

\bibitem{myers90}
R.~H. Myers, \emph{Classical and modern regression with applications}.\hskip
  1em plus 0.5em minus 0.4em\relax Duxbury Press, 1990, vol.~1.

\bibitem{MASS}
\BIBentryALTinterwordspacing
W.~N. Venables and B.~D. Ripley, \emph{Modern Applied Statistics with S},
  4th~ed.\hskip 1em plus 0.5em minus 0.4em\relax New York: Springer, 2002, iSBN
  0-387-95457-0. R MASS package Available at:
  https://cran.r-project.org/web/packages/MASS/index.html. Version 7.3.50.
  [Online]. Available: \url{http://www.stats.ox.ac.uk/pub/MASS4}
\BIBentrySTDinterwordspacing

\bibitem{kreyszig78}
E.~Kreyszig, \emph{Introductory functional analysis with applications}.\hskip
  1em plus 0.5em minus 0.4em\relax Wiley New York, 1978, vol.~1.

\end{thebibliography}

\newpage

\appendices

{\color{black}
\section{Calculating the $\bb_\lambda$-solutions from the SVD}
\label{Simplifications}
The full SVD of $\bX = \bU\bS\bV^\prime$ yields $\bV\bV^\prime =\bI_p$ and $\bX^\prime\bX = \bV\bS^\prime\bS\bV^\prime$. The right singular vectors $\bV$ of $\bX$ are obviously eigenvectors for both $\bX^\prime\bX$ and 
\begin{equation}\label{XtXlambdaI}\bX_{\lambda}^\prime\bX_{\lambda}=(\bX^\prime\bX+\lambda \bI_p)=\bV(\bS^\prime\bS+\lambda\bI_p)\bV^\prime,\end{equation}
and their corresponding eigenvalues are given by the diagonals of ${\bS}^\prime{\bS}$ and ${\bS}^\prime{\bS}+\lambda\bI_p$, respectively. The inverse 
matrix
$(\bX^\prime\bX+\lambda\bI_p)^{-1}=\bV(\bS^\prime\bS+\lambda\bI_p)^{-1}\bV^\prime$,
and the expression (\ref{bRR2}) for the TR-regression coefficients of a problem on standard form therefore simplifies \cite{hastie09} to
\begin{equation}\label{bRR3}
\begin{split}
\bb_{\lambda} & =\bV(\bS^\prime\bS+\lambda\bI_p)^{-1}\bV^\prime\bV\bS\bU^\prime\by \\
& =\bV(\bS^\prime\bS+\lambda\bI_p)^{-1}\bS\bU^\prime\by.
\end{split}
\end{equation}
In the following, we assume that $\bX$ has full rank, i.e., $r = rank(\bX) = \min(n,p)$. Then there are exactly $r$ non-zero rows in the $\bS$-factor of $\bb_{\lambda}$, and the zero rows of $\bS$ cancel both the associated columns in $\bV(\bS^\prime\bS+\lambda\bI_p)^{-1}$ and rows in $\bU^\prime$.
{By considering the compact SVD of} $\bX=\bU_r\bS_r\bV_r^\prime$ (the vanishing dimensions associated with the singular value $0$ are omitted from the factorisation), the expression (\ref{bRR3}) for the regression coefficients $\bb_{\lambda}$  simplifies to
\begin{equation}\label{bRRSVD}
\begin{split}
\bb_{\lambda} &=\bV_r(\bS_r^2+\lambda\bI_r)^{-1}\bS_r\bU_r^\prime\by \\
&=\bV_r(\bS_r+\lambda\bS_r^{-1})^{-1}\bU_r^\prime\by=\bV_r\bc_\lambda, 
\end{split}
\end{equation}
where the coordinate vectors
$\bc_\lambda=(\bS_r+\lambda\bS_r^{-1})^{-1}\bU_r^\prime\by=[c_{\lambda,1}\ ...\ c_{\lambda,r}]^\prime\in \R^r$ has the scalar entries 
\begin{equation}\label{ci}
c_{\lambda,j} = \frac{\bu_j^\prime\by}{s_j+\lambda/s_j}, \text{ for } 1\leq j\leq r. 
\end{equation}
}

{\color{black}

\section{A formula for the segmented cross-validation residuals in linear least squares regression}\label{fastSegCV}

The Sherman--Morrison--Woodbury updating formula for matrix inversion \cite{householder65} says that 
\begin{equation} \label{eq:Woodbury}
({\bA + UCV})^{-1} = \bA^{-1} - \bA^{-1}\bU(\bC^{-1} + \bV \bA^{-1} \bU)^{-1} \bV \bA^{-1},
\end{equation}
where $\bA$, $\bU$, $\bC$ and $\bV$ are conformable matrices ($\bA$ is $p\times p$, $\bC$ is $k\times k$, $\bU$ is $p\times k$, and $\bV$ is $k\times p$). The matrix identity (\ref{eq:Woodbury}) means that the inverse of the rank-$k$ modification of $\bA$ on the left-hand side can be obtained from a rank-$k$ modification of $\bA^{-1}$ that includes inversion of two rank $k$ matrices.
\\ \ \\
\noindent In the following we will use the notation
\begin{equation} \label{eq:Definintions}
\begin{aligned}
& \bA = \bX ' \bX,  \ \ \bU = \bX_{cv}',\ \ \bV = \bX_{cv},\ \ \\
&\bC = -\bI_{k} \text{ (the negative $k\times k$ identity matrix)},
\end{aligned}
\end{equation}
where the matrix $\bX_{cv}$ denotes a cross-validation sample segment obtained by selecting some $k$ rows from the full rank data matrix $\bX$. Moreover, the vector $\by_{cv}$ denotes the corresponding selection of entries from the response vector $\by$. Finally, let $(\bX_{(cv)},\by_{(cv)})$ denote the remaining rows of the full dataset $(\bX,\by)$ that are not contained in the sample segment $(\bX_{cv},\by_{cv})$, where we assume that also $\bX_{(cv)}$ has full rank.
\\ \ \\
\noindent
\textbf{Lemma \label{Lemma1}}\\
Let $\bM = \bX_{cv}(\bX' \bX )^{-1}$ and $\bH_{cv}=\bM\bX_{cv}'=\bX_{cv}(\bX' \bX )^{-1}\bX_{cv}'$. In the above notation, the following identity holds:
\begin{equation} 
\bX_{cv} (\bX_{(cv)}' \bX_{(cv)})^{-1} = [\bI_k - \bH_{cv}]^{-1}\bM. 
\end{equation}
\textbf{Proof:}\\
\noindent By substitutions according to the identities from (\ref{eq:Definintions}) into (\ref{eq:Woodbury}), we have:
{\small
\begin{equation} \label{eq:plugin}
\begin{split}
& (\bX_{(cv)}' \bX_{(cv)})^{-1} \\
&= (\bX' \bX - \bX_{cv}' \bX_{cv})^{-1} = (\bX' \bX - \bX_{cv}' \bI_{k} \bX_{cv})^{-1} \\
&= (\bX' \bX )^{-1} + (\bX' \bX )^{-1} \bX_{cv}' [\bI_{k} - \bX_{cv} (\bX' \bX)^{-1} \bX_{cv}']^{-1} \bX_{cv} (\bX' \bX)^{-1} \\
&= (\bX' \bX )^{-1} + \bM' [\bI_{k} - \bH_{cv}]^{-1} \bM.
\end{split}
\end{equation}
}

\noindent Multiplication of Equation (\ref{eq:plugin}) from the left by $\bX_{cv}$ yields:
\begin{equation} \label{eq:multiply}
\begin{split}
&\bX_{cv} (\bX_{(cv)}' \bX_{(cv)})^{-1} \\
&= \bM + \bH_{cv} [\bI_{k} - \bH_{cv}]^{-1}\bM \\
&= [\bI_{k} - \bH_{cv}][\bI_{k} - \bH_{cv}]^{-1}\bM + \bH_{cv} [\bI_{k} - \bH_{cv}]^{-1}\bM \\ 
&= [\bI_k - \bH_{cv}]^{-1}\bM. 
\end{split}
\end{equation}\nolinebreak
$\blacksquare $
\\ \ \\
\noindent 
By noting that 
\begin{equation} \label{Xty}
\bX_{(cv)}' \by_{(cv)}=(\bX' \by - \bX_{cv}' \by_{cv}), 
\end{equation} 
we are in the position to prove the following result for the prediction residuals of segmented cross-validation (SegCV): 
\\ \ \\
\textbf{Theorem (SegCV)\label{TmSegCV}}
\\
The prediction residuals $\br_{(cv)}=\by_{cv} - \bX_{cv} \hat{\boldsymbol \beta}_{(cv)}$ of the cross-validation sample segment $(\bX_{cv},\by_{cv})$ where $\hat{\boldsymbol \beta}_{(cv)}=(\bX_{(cv)}' \bX_{(cv)})^{-1}\bX_{(cv)}' \by_{(cv)}$, can alternatively be obtained by a linear transformation of the associated fitted residuals $\br_{cv} = (\by_{cv}-\bX_{cv}\hat{\boldsymbol \beta})$ from the full model 
$\hat{\boldsymbol \beta} = (\bX' \bX)^{-1}\bX' \by$ as follows:
\begin{equation} \label{eq:residuals}
\br_{(cv)} = [\bI_{k} - \bH_{cv}]^{-1} \br_{cv}. 
\end{equation}
\noindent
\textbf{Proof:}\\
{\small
\begin{equation} 
\begin{split}
&\br_{(cv)} \\
&= \by_{cv} - \bX_{cv} \hat{\boldsymbol \beta}_{(cv)} \\
&= \by_{cv} - \underbrace{\bX_{cv} (\bX_{(cv)}' \bX_{(cv)})^{-1}}_{\text{(\ref{eq:multiply})}} \underbrace{\bX_{(cv)}' \by_{(cv)}}_{\text{(\ref{Xty})}}\\
&= [\bI_{k} - \bH_{cv}]^{-1} [\bI_{k} - \bH_{cv}] \by_{cv} - [\bI_{k} - \bH_{cv}]^{-1} \bM[\bX' \by - \bX_{cv}' \by_{cv}]\\
&= [\bI_{k} - \bH_{cv}]^{-1} [\by_{cv} - \bH_{cv} \by_{cv} -\ \ \ \ \ \ \  \bM\bX' \by \ \ \ \ \ \  + \ \ \bM\bX_{cv}'\by_{cv}]\\
&= [\bI_{k} - \bH_{cv}]^{-1} [\by_{cv} - \bH_{cv} \by_{cv} - \bX_{cv} \underbrace{(\bX' \bX)^{-1}\bX' \by}_{\hat{\boldsymbol \beta}} + \ \ \ \ \ \bH_{cv}\by_{cv}]\\
&= [\bI_{k} - \bH_{cv}]^{-1}(\by_{cv}-\bX_{cv}\hat{\boldsymbol \beta})\\ 
&= [\bI_{k} - \bH_{cv}]^{-1} \br_{cv}. 
\end{split}
\end{equation}}
$\blacksquare $
\\ \ \\
Equation (\ref{eq:residuals}) shows that we can calculate the prediction residuals for a cross-validation sample segment of size $k$ at the cost of inverting the $k\times k$ matrix $[\bI_{k} - \bH_{cv}]$ followed by a matrix-vector multiplication with the fitted residuals $\br_{cv}$.

The case of $k = 1$ corresponds to leave-one-out cross-validation (LooCV) where the prediction residual calculations reduce to scalar operations:
\\ \ \\
\textbf{Corollary (LooCV)\label{CoLooCV}}
\\*
The prediction residual $r_{(i)}$ when holding out the $i$-th sample $(\bx_i, y_i)$ from the modelling is 
\begin{equation} \label{LooCVres}
r_{(i)} = r_i/(1-h_i),
\end{equation}
where $r_i = y_i - \bx_i\hat{\boldsymbol \beta}$ is the fitted residual and $h_i = \bx_i(\bX'\bX)^{-1}\bx_i'$.
$\blacksquare $
\\ \ \\
Here, $h_i$ is the $i$-th diagonal element of the projection matrix $\bH=\bX(\bX'\bX)^{-1}\bX'$ (projection onto the column space of $\bX$).
}

\section{The Tikhonov $L_2$-regularisation framework} \label{secL2}
Tikhonov \cite{tikhonov63} noted that it is straightforward to generalise the above $L_2$  regularisation of $\bb$ to more specialised types of regularisation through a corresponding regularisation matrix $\bL$. These cases are expressed in terms of identifying the minimising solution of the bi-objective least squares problem
\begin{equation}\label{GTR}
\begin{split}
RSS_{\bL,\lambda}(\bb)&=\|\bX\bb-\by\|^2+\lambda\|\bL\bb-\bO\|^2 \\
&=\|\bX\bb-\by\|^2+\lambda\|\bL\bb\|^2,
\end{split}
\end{equation}
for some fixed $\lambda>0$.
The minimisation of Equation (\ref{GTR}) with respect to $\bb$ can be obtained by considering the augmented data $\bX_{\bL,\lambda}=\left[%
\begin{array}{c}
  \bX \\
  \sqrt{\lambda}\bL \\
\end{array}%
\right] \text{ and } \by_{0}=\left[%
\begin{array}{c}
  \by \\
\bO \\
\end{array}%
\right]
$, and solving the normal equations 
\begin{equation}\label{TRnormaleq} 
\bX_{\bL,\lambda}^\prime\bX_{\bL,\lambda}\bb=\bX_{\bL,\lambda}^\prime\by_0\Rightarrow (\bX^\prime\bX+\lambda\bL^\prime\bL)\bb=\bX^\prime\by
\end{equation}
associated with the OLS problem $\bX_{\bL,\lambda}\bb=\by_0$.

To avoid technical distractions we will in the following restrict our attention to the cases of square and non-singular regularisation matrices $\bL$
(even for situations where a non-square regularisation matrix is the immediate choice, a non-singular $(p\times p)$-alternative that provides a good approximation is often available).
By defining $\tilde{\bX}=\bX\bL^{-1}$, the solution of the OLS problem in (\ref{TRnormaleq}) is equivalent to finding the unique OLS-solution $\bm{\beta}_{\lambda}$ of the transformed problem
$\tilde{\bX}_{\lambda}\bm{\beta}=\by_0$,
where $\tilde{\bX}_{\lambda}={\bX}_{\bL,\lambda}{\bL}^{-1}= \left[%
\begin{array}{c}
  \tilde{\bX} \\
  \sqrt{\lambda}{\bI} \\
\end{array}%
\right]$ and $\bm{\beta}={\bL}{\bb}$. 
The associated expression  minimised by $\bm{\beta}_{\lambda}$ is 
\begin{equation}\label{GTReqv}
\|\tilde{\bX}\bm{\beta}-\by\|^2+\lambda\|\bm{\beta}\|^2,
\end{equation}
i.e., in the standard form (\ref{RR}), and the minimising solution $\bb_\lambda$ of the original problem (\ref{GTR}) is obtained by 
\begin{equation}\label{backtransf}\bb_\lambda={\bL}^{-1}\bm{\beta}_{\lambda}.
\end{equation}
Among the many useful choices for the regularisation matrix ${\bL}$ are the following:
\begin{enumerate}
\item diagonal scaling (e.g., the standardisation of variables often advised for RR applications):\\ $\bL_{std}=\left[%
\begin{array}{cccc}
  \hat{\sigma}_1 &  &  &\\
  & \hat{\sigma}_2 &  & \\
  & & \ddots &          \\
  & & & \hat{\sigma}_p
\end{array}%
\right]$,\\
where $\hat{\sigma}_i$ approximates the standard deviation of the $i$-th variable $(1\leq i\leq p)$.\\ \ \\
\item a (full) rank $p$ discrete 1. derivative approximation:\\ $\bL_{1}=\left[%
\begin{array}{ccccc}
  1 & -1 & &  & \\
  & 1 & -1 &  & \\
  & & \ddots & \ddots &\\
  & & & 1 & -1 \\
  \sqrt{\epsilon} c_1 & \sqrt{\epsilon} c_1 & \dots & \sqrt{\epsilon} c_1 & \sqrt{\epsilon}c_1 \\
\end{array}%
\right]$.\\ \ \\
\item a (full) rank $p$ discrete 2. derivative approximation:\\  $\bL_{2}=$\\
{\footnotesize \noindent$\left[
\arraycolsep=1.8pt
\begin{array}{cccccc}
  1 & -2 & 1 &  &  \\
  & 1 & -2 & 1 &  &\\
  & & \ddots & \ddots & \ddots &\\
  & & & 1 & -2 & 1\\
\sqrt{\epsilon} c_1 & \sqrt{\epsilon} c_1 & \dots & \sqrt{\epsilon} c_1 & \sqrt{\epsilon} c_1 & \sqrt{\epsilon} c_1 \\
-\sqrt{\epsilon} c_2 \frac{p}{2} & -\sqrt{\epsilon} c_2 \frac{p-1}{2} & \; \; \; \; \dots \; \; \; \; & \; \; \; \; \dots \; \; \; \; &  \sqrt{\epsilon} c_2 \frac{p-1}{2} & \sqrt{\epsilon} c_2 \frac{p}{2} \\
\end{array}%
\right]$.}
\end{enumerate}
\vspace{2mm}

\noindent The alternatives $\bL_1$ and $\bL_2$ are relevant for problems where the $\bX$-data are associated with discretised (uniform) sampling of continuous signals so that some smoothness in the solution candidates $\bb_\lambda$ is reasonable.
The two last rows in $\bL_2$ (and the last row in $\bL_1$) above are scaled versions of the discretised and normalised Legendre polynomials \cite{kreyszig78} of order $0$ and $1$, respectively ($c_1$ and $c_2$ represent the normalisation constants, and $\epsilon>0$ is a scaling factor to be commented on below). It should be noted that both these row vectors are orthogonal to all the above rows in the derivative matrices where they appear. 

The main purpose of the included Legendre vectors in these regularisation matrices is to ensure 
full rank of the regularisation matrices.
Appropriate regularisation of the solutions $\bb_\lambda$ may be obtained by choosing the fixed scaling factor $\epsilon>0$ to be 
\begin{itemize}
\item either sufficiently large to make $\bb_\lambda$ practically orthogonal to the subspace spanned by the Legendre vectors, or 
\item sufficiently small to inhibit any notable penalisation with respect to the same Legendre vectors.
\end{itemize}
The choice of $\epsilon$ in the last case can therefore not be made arbitrarily small in practice. It must be chosen large enough to avoid numerical difficulties in the computations of $\tilde{\bX}$ and $\bb_\lambda$. Alternative (non-invertible) differentiation matrix candidates taking various boundary conditions into account are described in \cite{hansen10}.

\section{Special situation for segment decomposition in VirCV }\label{app:VirCVsituations}
In the following we will examine the proposed VirCV strategy more closely for three different situations: 
\begin{enumerate}
\item[a)] Segments of identical rows. 
\item[b)] Segments of collinear rows.
\item[c)] The general case (segments with no particular structure in the rows). 
\end{enumerate}
\textbf{Identical rows:}\\ Let us assume that all the rows of a segment $\bX_i,\ (1\leq i\leq K)$ are identical. In this particular case, the $PRESS$-function associated with the VirCV is identical to the $PRESS$-function obtained by the SegCV. 

The identity can be derived by noting that the left-multiplication of the left- and right-hand sides of a linear system by an orthogonal matrix affects neither the least squares solution nor the norm of the associated residual vector. Consequently, the SegCV strategy applied to the two systems (\ref{boeq1}) and (\ref{boeq2}) will result in identical $PRESS$-functions. 
With all rows within each segment $\bX_k\in \mathbb{R}^{n_k\times p}$ being identical to its first row (denoted $\bx_{k,1}$) of the segment, it is straightforward to verify that $\bX_k$ has only one non-zero singular value $s_{k,1}=\sqrt{\bx_{k,1}\bx_{k,1}^\prime n_k}$ and the corresponding left- and right singular vectors are 
\begin{equation}\bu_{k,1}= \frac{1}{\sqrt{n_k}}\left[\begin{array}{c}
  1 \\
  1\\
  : \\
  1
\end{array}%
\right]\in \mathbb{R}^{n_k} \text{ and } \bv_{k,1}=\frac{1}{\sqrt{\bx_{k,1}\bx_{k,1}^\prime}}\bx_{k,1}^\prime\in \mathbb{R}^p.
\end{equation}
\noindent
By the orthogonality requirements of the SVD, any other left singular vector $\bu$ must satisfy $\bu^\prime\bu_{k,1}=0$. Consequently

\begin{equation}
\bU_k^\prime\bX_k = \left[\begin{array}{c}
  \sqrt{n_k}\bx_{k,1} \\
  \bm{0}\\
  : \\
  \bm{0}
\end{array}%
\right]
\text{ and  }
\bU_k^\prime\bm{1} = \left[\begin{array}{c}
\sqrt{n_k} \\
  0\\
  : \\
  0
\end{array}%
\right],
\end{equation}
 meaning that there will be only one non-zero row in each segment on the left-hand side of the $\tilde{\bT}$-transformed system (\ref{boeq2}).
It is therefore sufficient to demonstrate that the $PRESS$-functions obtained from applying the SegCV and the LooCV to the system in \eqref{boeq2} are equal: 
Clearly, for any row containing just zeros in the left-hand side of \eqref{boeq2} the prediction based on it is trivially identical to $0$ (zero) for either of the cross-validation strategies (regardless of the regression coefficients).
Because such zero rows do not contribute to the calculation of the regression coefficients, we are forced to conclude that the regression coefficients obtained by holding out the (only) non-zero row of a segment must be equal to the regression coefficients obtained from holding out the entire segment. Thus the predicted values for the non-zero row in each segment must also be identical for both cross-validation strategies, and we can 
conclude that the $PRESS$ functions obtained by the SegCV- and the VirCV strategies must be identical.

\vspace{3mm}
\noindent
\textbf{Collinear (proportional) rows:}\\  
One might expect the same result to hold when the rows within a segment are proportional. This is however not the case with the modelling strategy described above. The reason for this is that the inclusion of a constant term will make each of the $K$ segments become a rank $2$ -- rather than a rank $1$ submatrix. 
With more than one non-zero row on the left-hand side in each segment the argument of the previous situation fails, 
and doing LooCV on the transformed data is no longer equivalent to 
doing SegCV on the original data. 
However, when omitting the constant term from the modelling, each of the $K$ segments has rank 1, and the SegCV and VirCV approaches will result in identical $PRESS(\lambda)$-functions. The rigorous explanation is similar to the argument given for the situation with identical rows.

\newpage
\section{TR Prototype MATLAB code}\label{trcode}
\begin{lstlisting}
function [press, bcoefs, b, lambda, H, U, s, V, GCV, ...
          L, idmin, rescv] = TregsLooCV(X, y, lambdas, type)
% ----------------------------------------------------------
% INPUTS:
% X       - Data matrix
% y       - Response vector
% lambdas - Vector of regularisation parameter values
% type    - Regularisation type (-1 for standardisation, 
%           0 for L2, 1 for 1st derivative regularisation, 
%           etc ...)
% ----------------------------------------------------------
% OUTPUTS:
% press   - PRESS-statistic for input lambdas
% bcoefs  - Regression coefficients for selected lambda 
%           (no constant term)
% b       - Regression coefficients for PRESS-minimal lambda 
%           (with constant term)
% lambda  - Value of lambda minimising the PRESS-statistic
% H       - Vector of leverage values for all values of lambda
% U, s, V - SVD of matrix
% GCV     - GCV-statistic for input lambdas
% L       - Regularisation matrix 
%           (empty for L2 regularisation)
% idmin   - Index of lambda value minimising 
%           the PRESS-statistic
% rescv   - LooCV-residuals
% ----------------------------------------------------------

[n,p] = size(X);
mX = mean(X); my = mean(y);
X = bsxfun(@minus,X,mX); y = y-my;

L = [];
% Create full rank discrete derivative matrix of order 'type'.
if type > 0 
    epsilon = 1e-14;
    L = diff([speye(p);sparse(type,p)],type);
    L(end-type+1:end,:) = sqrt(epsilon)*Plegendre(type-1,p);
elseif type < 0 % Create variable standardisation matrix.
    L = spdiags(std(X)',0,p,p);
end
if type ~= 0, X = X/L; end

[U, S, V] = svd(X,'econ'); s = diag(S);
denom     = bsxfun(@plus,s,bsxfun(@rdivide,lambdas,s));
bcoefs    = V*bsxfun(@rdivide,(U'*y),denom);
H         = (U.^2)*bsxfun(@rdivide,s,denom)+1/n;
resid     = bsxfun(@minus,y, ...
                   U*bsxfun(@rdivide,s.*(U'*y),denom));
rescv     = bsxfun(@rdivide,resid,(1-H));
press     = sum(rescv.^2)';
GCV       = (sum(resid.^2)./mean(1-H).^2)';

% Finding press-minimal model and corresponding 
% regression coefficients:
[~,idmin] = min(press); lambda = lambdas(idmin); 
h = H(:,idmin);
if type  ~= 0, bcoefs = L\bcoefs; end
% Constant term
b         = [my-mX*bcoefs(:,idmin); bcoefs(:,idmin)]; 
end

function Q = Plegendre(d,p)
P = ones(p,d+1);
x = (-1:2/(p-1):1)';
for k = 1:d
    P(:,k+1) = x.^k;
end
[Q,~] = qr(P,0);
Q = Q';
end
\end{lstlisting}

\newpage
\section{SegCV Prototype MATLAB code}\label{SegCVcode}
\begin{lstlisting}
function [press, bcoefs, b, lambda, H, U, s, V, GCV, L, ...
          idmin, rescv] = TregsSegCV(X, y, lambdas, ...
                                     type, cv)
% ----------------------------------------------------------
% INPUTS:
% X       - Data matrix
% y       - Response vector
% lambdas - Vector of regularisation parameter values
% type    - Regularisation type (-1 for standardisation, 
%           0 for L2, 1 for 1st derivative regularisation, 
%           etc ...)
% cv      - Vector of cross-validation segments 
%           (integers from 1 to #segments, length n)
% ----------------------------------------------------------
% OUTPUTS:
% press   - PRESS-statistic for input lambdas
% bcoefs  - Regression coefficients for selected lambda 
%           (no constant term)
% b       - Regression coefficients for PRESS-minimal lambda 
%           (with constant term)
% lambda  - Value of lambda minimising the PRESS-statistic
% H       - Vector of leverage values for all values of lambda
% U, s, V - SVD of matrix
% GCV     - GCV-statistic for input lambdas
% L       - Regularisation matrix 
%           (empty for L2 regularisation)
% idmin   - Index of lambda value minimising the 
%           PRESS-statistic
% rescv   - LooCV-residuals
% ----------------------------------------------------------

[n,p] = size(X);
nseg = max(cv); % Number of CV-segments
nlambda = length(lambdas);
mX = mean(X); my = mean(y);
X = bsxfun(@minus,X,mX); y = y-my;

L = [];
% Create full rank discrete derivative matrix of order 'type'.
if type > 0 
    epsilon = 1e-14;
    L = diff([speye(p);sparse(type,p)],type);
    L(end-type+1:end,:) = sqrt(epsilon)*Plegendre(type-1,p);
elseif type < 0 % Create variable standardisation matrix.
    L = spdiags(std(X)',0,p,p);
end
if type ~= 0, X = X/L; end

[U, S, V] = svd(X,'econ'); s = diag(S);
denom     = bsxfun(@plus,s,bsxfun(@rdivide,lambdas,s));
bcoefs    = V*bsxfun(@rdivide,(U'*y),denom);
H         = (U.^2)*bsxfun(@rdivide,s,denom)+1/n;
resid     = bsxfun(@minus,y, 
                   U*bsxfun(@rdivide,s.*(U'*y),denom));
rescv     = zeros(n,nlambda);
sdenom  = sqrt(bsxfun(@rdivide,s,denom))'; 
for seg = 1:nseg
    Useg = U(cv==seg,:); I = eye(size(Useg,1)) - 1/n;
    for k = 1:nlambda
        Uk = Useg.*sdenom(k,:);
        % Exact CV using H correction
        rescv(cv==seg,k) = (I - Uk*Uk')\resid(cv==seg,k); 
    end
end
press     = sum(rescv.^2)';
GCV       = (sum(resid.^2)./mean(1-H).^2)';

% Finding press-minimal model and corresponding 
% regression coefficients:
[~,idmin] = min(press); lambda = lambdas(idmin); 
h = H(:,idmin);
if type  ~= 0, bcoefs = L\bcoefs; end
% Constant term
b         = [my-mX*bcoefs(:,idmin); bcoefs(:,idmin)]; 
end
\end{lstlisting}

\newpage
\section{VirCV Prototype MATLAB code}\label{VirCVcode}
\begin{lstlisting}
function [press, bcoefs, b, lambda, H, U, s, V, GCV, L, ...
          idmin, rescv, Usegments] = TregsVirCV(X, y, ...
                                     lambdas, type, segments)
% ----------------------------------------------------------
% INPUTS:
% X         - Data matrix
% y         - Response vector
% lambdas   - Vector of regularisation parameter values
% type      - Regularisation type (-1 for standardisation, 
%             0 for L2, 1 for 1st derivative regularisation,
%             etc ...)
% segments  - List of integers identifying 
%             cross-validation segments
% ----------------------------------------------------------
% OUTPUTS:
% press     - PRESS-statistic for input lambdas
% bcoefs    - Regression coefficients for selected lambda 
%             (no constant term)
% b         - Regression coefficients for PRESS-minimal lambda 
%             (with constant term)
% lambda    - Value of lambda minimising the PRESS-statistic
% H         - Vector of leverage values for all 
%             values of lambda
% U, s, V   - SVD of matrix
% GCV       - GCV-statistic for input lambdas
% L         - Regularisation matrix 
%             (empty for L2 regularisation)
% idmin     - Index of lambda value minimising the 
%             PRESS-statistic
% rescv     - LooCV-residuals
% Usegments - Sparse matrix representing the orthogonal 
%             transformations used in the VirCV
% ----------------------------------------------------------

% Finding orthogonal transformation and the modification 
% to the leverage correction:
Usegments = segmentORTH(X, segments);
bs = (sum(Usegments,1).^2)';

[n,p] = size(X);
mX = mean(X); my = mean(y);
X = bsxfun(@minus,X,mX); y = y-my;

% Transforming data:
X = Usegments'*X; y = Usegments'*y;

L = [];
if type > 0
    epsilon = 1e-14;
    L = diff([speye(p);sparse(type,p)],type);
    P = Plegendre(type-1,p);
    L(end-type+1:end,:) = sqrt(epsilon)*P;
elseif type < 0
    L = spdiags(std(X)',0,p,p);
end

if type ~= 0, X = X/L; end

[U, S, V]             = svd(X,'econ'); s = diag(S);
s_plus_lambdas_over_s = bsxfun(@plus,s, ...
                               bsxfun(@rdivide,lambdas,s));

H      = bsxfun(@plus, (U.^2)*bsxfun(@ldivide,...
                s_plus_lambdas_over_s, s), bs/n);
bcoefs = V*bsxfun(@ldivide,s_plus_lambdas_over_s,(U'*y));
res    = bsxfun(@minus,y,X*bcoefs);
rescv  = bsxfun(@rdivide,res,(1-H));
press  = sum(rescv.^2)';
GCV    = sum(bsxfun(@rdivide,res,mean(1-H)).^2)';

if type ~= 0, bcoefs = L\bcoefs; end

% Finding press-minimal model and corresponding 
% regression coefficients:
[~,idmin] = min(press); lambda = lambdas(idmin); 
h = H(:,idmin);
if type  ~= 0, bcoefs = L\bcoefs; end
% Constant term
b         = [my-mX*bcoefs(:,idmin); bcoefs(:,idmin)]; 

end

function U = segmentORTH(X, segments)
n 		  = size(X,1);
nsegments = max(segments);
U 		  = sparse(n,n);
for k = 1:nsegments
    ind 			= find(segments==k);
    [U(ind, ind),~] = svd(X(ind,:),'econ');
end
end

function Q = Plegendre(d,p)
P = ones(p,d+1);
x = (-1:2/(p-1):1)';
for k = 1:d
    P(:,k+1) = x.^k;
end
[Q,~] = qr(P,0);
Q = Q';
end
\end{lstlisting}

\section{Computational complexity of the fast LooCV}\label{floploocv}
For a more precise description of the computational complexity involved in calculating the fast LooCV, an approximate count of the floating point operations (flop) 
is required. According to Bj\"orck \cite{bjorck16}, an approximate flop count for finding the reduced SVD (using a QR-SVD algorithm with Golub--Kahan--Householder bidiagonalisation) of a $(n\times p)$-matrix is $12pn^2+(16/3)n^3$ when assuming $p\ge n$. The remaining computations consist of centring, calculating the $PRESS$ values, and calculating the $PRESS$-minimal regression coefficients for every response. 
With $q$ different responses, the approximate flop count for these computations is given by:
\begin{equation}
\begin{aligned}
&(3np+3nq+nr+2nrq-q+2prq+pq)\\
&+n_\lambda(3r+2nr+2nrq+qr+4nq),
\end{aligned}
\end{equation}
\noindent
where $n_\lambda$ denotes the number of different candidate regularisation parameter values.
For $p\ge n$, the computations needed to evaluate the $PRESS(\lambda)$-function for one additional regularisation parameter is of the order $\mathcal{O}(qn^2)$, and in particular the additional computations are independent of the number $(p)$ of measured features. This makes the fast LooCV highly useful also for 
problems where the number of features is even larger than the number of samples. To calculate the cost of finding the corresponding $GCV(\lambda)$-values as well as $GCV$-minimal regression coefficients one should add $5nn_\lambda q-q+q(2pr+p)$ to the above flop count. Note that the choice of regularisation matrix $\bL$ matters here, and
for $\bL\ne \bI$ there are additional calculations (see Section \ref{secL2}) that must be taken into account. The exact number of flops associated with these additional calculations will depend on the sparsity structure of $\bL$ and to what extent that sparsity can be utilised in the required calculations.

\bgroup
\def\arraystretch{1.3} 
\begin{table}[!htb] 
\begin{tabular}{|c|c|} \hline
Par. sel. method & Approx. flops for SVD(s) \\ \hline
LooCV/GCV & $12pn^2+\frac{16}{3}\cdot n^3$ \\ \hline
VirCV & $12pn^2+\frac{16}{3}\cdot n^3$ + $K\cdot\left(12p\cdot B_{ss}^2+\frac{16}{3}\cdot B_{ss}^3\right)$ \\  \hline
SegCV & $K\cdot\left(12p\cdot (n-B_{ss})^2+\frac{16}{3}\cdot (n-B_{ss})^3\right)$ \\ \hline
\end{tabular}
\caption{\textit{Approximate flop counts for the required SVD(s) in the different parameter selection methods when assuming $p\ge n$.}}
\label{tab:SVDflops}
\end{table}
\egroup

\newpage

\section{Computational savings of the VirCV compared to the SegCV} \label{flopVirCV}
To assess the computational savings of the VirCV over the SegCV, flop count approximations for the associated $PRESS$-values 
must be compared. (We only consider the situation involving $L_2$ regularisation, i.e. the identity matrix $\bI$ acting as the regularisation matrix.)
Let $K$ denote the number of segments, and assume for simplicity that the various segment sizes are all bounded from above by the constant $B_{ss}$. The approximate number of flops required for the SVDs for the different parameter selection methods when using the entire dataset for training are given by the formulas in Table \ref{tab:SVDflops} (using the approximate flop count for the SVD given in \cite{bjorck16}). 
{\color{black}The Table shows that the size of (all but one of) the required SVDs for the VirCV are much smaller than for the SegCV}
(assuming the size of each segment is much smaller than the total number of samples, which is obviously the case in most real applications). This is primarily what makes the VirCV superior to 
the SegCV in terms of computational efficiency.

If the block diagonal structure of the transformation matrix $T$ is utilised, the matrix multiplication part of the orthogonal transformation  \eqref{boeq2} for the VirCV requires approximately

\begin{equation}
2B_{ss}(B_{ss}-1)+K\cdot B_{ss}\cdot p(2B_{ss}-1)+q\cdot B_{ss}(2B_{ss}-1)
\end{equation}
\noindent
flops. For keeping track of the remaining computations needed for the VirCV we can use the flop count approximations in Section \ref{floploocv}, as the flop count for the VirCV and the LooCV will be identical after applying the orthogonal transformation required for the VirCV. The approximate flop count of the remaining computations for the SegCV is given by
\begin{equation}
\begin{aligned}
&2K\cdot B_{ss}(q+p)-q+r_{train}\cdot q(2B_{ss}-1)\\
&+q\cdot n_\lambda\cdot K[3r_{train}+2p\cdot r_{train}+p+2p\cdot n_{test}+3n_{test}]
\end{aligned}
\end{equation}
\noindent
where $r_{train}=\min(n_{train},p)$ and $n_{train}$ is the number of samples in the training set.

{\color{black}Although the main computational cost with model validation is with the initial SVD(s) there will also be an additional computational cost for each candidate regularisation parameter value for which we want to validate the model. Consider the case $p>n$ of most interest for the present work (the number of features is greater than the number of samples). From the above reasoning, we observe that 
when considering additional regularisation parameter values, the SegCV flop count depends on the number of features $p$ for each candidate value. The above flop count for the VirCV and the LooCV flop count in Appendix \ref{floploocv} shows that this is not the case for the VirCV. When $p$ is very large it might therefore be computationally inefficient (or even infeasible) to validate models for a large number of regularisation parameter values based on the SegCV. Clearly, the VirCV is the method of choice among the two in such cases.}

\begin{IEEEbiography}[{\includegraphics[width=1in,height=1.25in,clip,keepaspectratio]{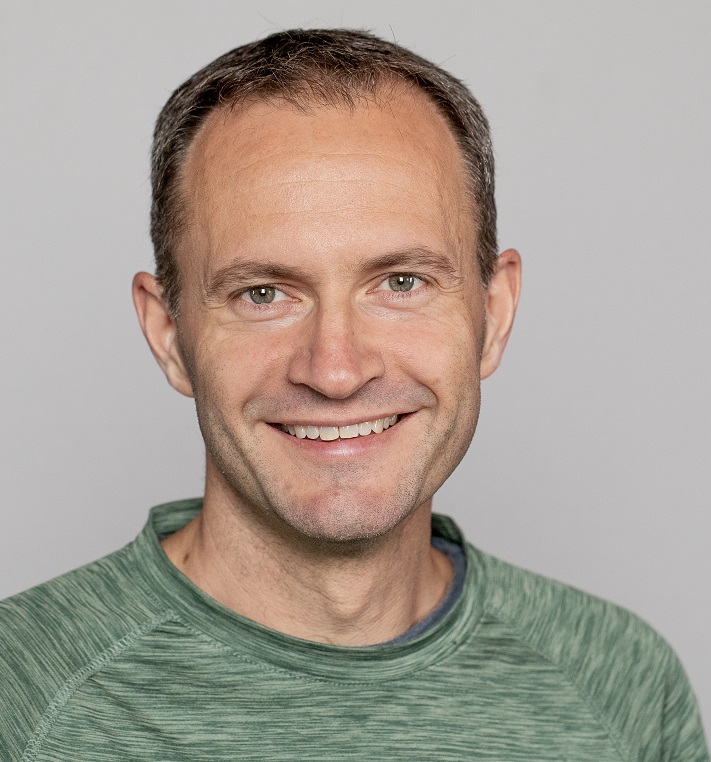}}]{Kristian Hovde Liland} is a professor in statistics at the Faculty of Science and Technology at the Norwegian University of Life Sciences. He received his PhD degree in applied statistics in 2010. His interests are mainly
within Multivariate Statistics and Machine Learning with a large spectrum of projects ranging from theoretic to applied and often with an emphasis on scientific programming.
\end{IEEEbiography}

\begin{IEEEbiography}[{\includegraphics[width=1in,height=1.25in,clip,keepaspectratio]{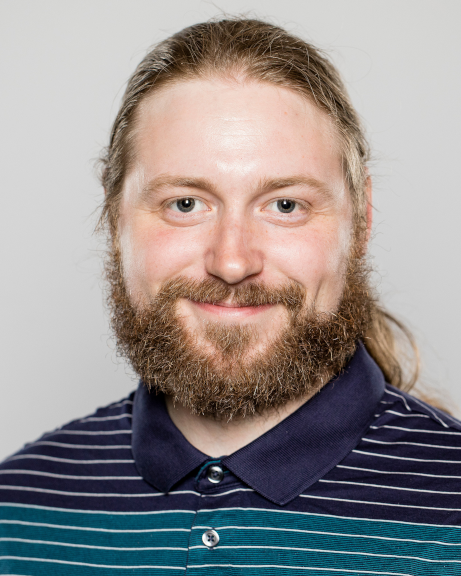}}]{Joakim Skogholt} received the M.Sc. degree in mathematics from the University of Warwick in 2012, and the Ph.D. degree in applied mathematics and data analysis from the Norwegian University of Life Sciences (NMBU) in 2019. He is currently employed as assistant professor at NMBU. His research interests include Machine Learning and Multivariate Statistics.
\end{IEEEbiography}

\begin{IEEEbiography}[{\includegraphics[width=1in,height=1.25in,clip,keepaspectratio]{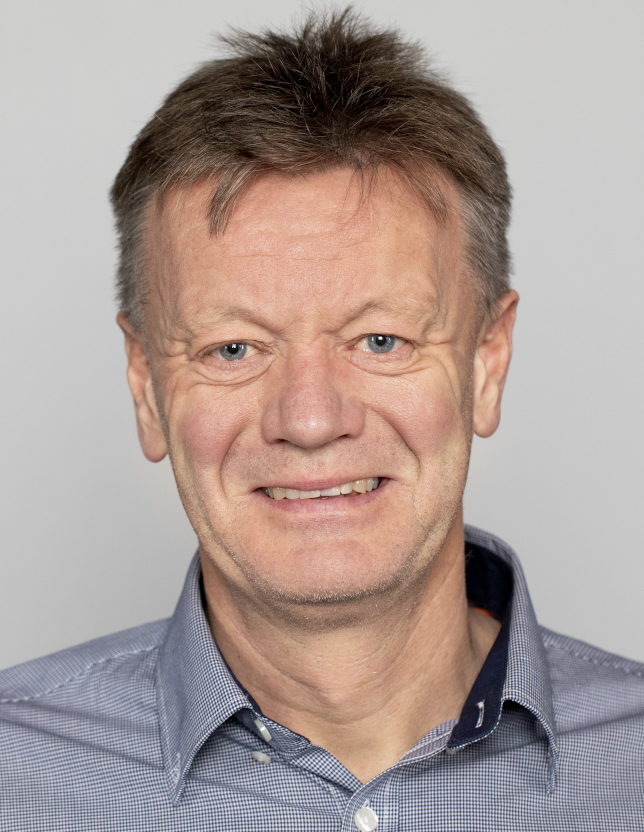}}]{Ulf Geir Indahl} is a Professor in
statistics at the Faculty of Science and Technology,
Norwegian University of Life Sciences. He
received his PhD degree in applied mathematics in
1998 from the University of Oslo. He is
interested in the development of new methodology
as well as applications ranging
from high-dimensional Multivariate Data Analysis
to Machine Learning and Deep Learning.
\end{IEEEbiography}

\EOD

\end{document}